\documentclass[aip,jcp, reprint, groupedaddress, floatfix]{revtex4-2}%
\pdfoutput=1
\usepackage{graphicx}
\usepackage{tabularx}
\usepackage{dcolumn}
\usepackage{longtable}
\usepackage{tensor}
\usepackage{color}
\usepackage{xcolor}
\usepackage{placeins}
\usepackage{bm}
\usepackage{amsmath}
\usepackage{amsfonts}
\usepackage{amssymb}
\usepackage{enumerate}
\providecommand{\U}[1]{\protect\rule{.1in}{.1in}}
\colorlet{RED}{red}

\graphicspath{{figures/}{C:/Users/Julien/Dropbox/Julien/Papers/integrators_local_control_theory/figures/}{C:/Users/Jiri/Dropbox/Papers/Chemistry_papers/2020/integrators_local_control_theory/figures/}}
\begin{document}
\title{Time-reversible and norm-conserving high-order integrators for the nonlinear
time-dependent Schr\"{o}dinger equation:\ Application to local control theory}
\author{Julien Roulet}
\email{julien.roulet@epfl.ch}
\author{Ji\v{r}\'{\i} Van\'{\i}\v{c}ek}
\email{jiri.vanicek@epfl.ch}
\affiliation{Laboratory of theoretical physical chemistry, Institut des sciences et
ing\'{e}nieries Chimiques, Ecole Polytechnique F\'{e}d\'{e}rale de Lausanne
(EPFL), Lausanne, Switzerland}
\date{\today}

\begin{abstract}
{ The explicit split-operator algorithm has been extensively used
for solving not only linear but also nonlinear time-dependent Schr\"{o}dinger
equations. When applied to the nonlinear Gross-Pitaevskii equation, the method
remains time-reversible, norm-conserving, and retains its second-order
accuracy in the time step. However, this algorithm is not suitable for all
types of nonlinear Schr\"{o}dinger equations. Indeed, we demonstrate that
local control theory, a technique for the quantum control of a molecular
state, translates into a nonlinear Schr\"{o}dinger equation with a more
general nonlinearity, for which the explicit split-operator algorithm loses
time reversibility and efficiency (because it has only first-order accuracy).
Similarly, {the trapezoidal rule (the Crank--Nicolson method)}, while
time-reversible, does not conserve the norm of the state propagated by a
nonlinear Schr\"{o}dinger equation. To overcome these issues, we present
high-order geometric integrators suitable for general time-dependent nonlinear
Schr\"{o}dinger equations and also applicable to nonseparable Hamiltonians.
These integrators, based on the symmetric compositions of the implicit
midpoint method, are both norm-conserving and time-reversible. The geometric
properties of the integrators are proven analytically and demonstrated
numerically on the local control of a two-dimensional model of retinal. For
highly accurate calculations, the higher-order integrators are more efficient.
For example, for a wavefunction error of $10^{-9}$, using the eighth-order
algorithm yields a $48$-fold speedup over the second-order implicit midpoint {method}
and { trapezoidal rule}, and $400000$-fold speedup over the explicit
split-operator algorithm.}

\end{abstract}
\maketitle




\section{{Introduction}}

 Nonlinear time-dependent Schr\"{o}dinger equations contain, by
definition, Hamiltonians that depend on the quantum state. Such
state-dependent effective Hamiltonians appear in many areas of physics and
chemistry. Examples include various nonlinear Schr\"{o}dinger equations
generated by the Dirac-Frenkel time-dependent variational principle,
\cite{book_Frenkel:1934, Dirac:1930, Broeckhove_Lathouwers:1988,
book_Lubich:2008} e.g., the equations of the multi-configurational
time-dependent Hartree method,\cite{Meyer_Cederbaum:1990,
Manthe_Cederbaum:1992, Beck_Jackle:2000} and some numerical methods such as
the short-iterative Lanczos algorithm.\cite{Lanczos:1950,
Leforestier_Kosloff:1991, Park_Light:1986} Probably the best known nonlinear
Schr\"{o}dinger equations, however, are approximate equations for
Bose-Einstein condensates,\cite{Anderson_Cornell:1995, Dalfovo_Stringari:1999}
in which the Hamiltonian depends on the probability density of the quantum
state. The dynamics of a Bose-Einstein condensate is often modeled by solving
the celebrated Gross-Pitaevskii equation with a cubic
nonlinearity.\cite{Gross:1961, Pitaevskii:1961, Carles:2002,
Carles_Sparber:2008,Minguzzi_Vignolo:2004} 

To solve this equation, several numerical schemes, such as the explicit
split-operator algorithm or the time and spatial finite difference methods
have been employed.\cite{Bao_Markowich:2003,Chang_Sun:1999} These methods are
of low accuracy (in time and/or space) and do not always preserve the
geometric properties of the exact solution.\cite{Antoine_Besse:2013} For
example, the Crank--Nicolson finite difference method is geometric but exhibits only second-order convergence with respect to the spatial
discretization. To remedy this, the explicit second-order split-operator
algorithm,\cite{Feit_Steiger:1982,Kosloff_Kosloff:1983, Kosloff_Kosloff:1983b,
book_Tannor:2007} commonly used for the linear time-dependent Schr\"{o}dinger
equation, is a great alternative, as it conserves, in some cases, the
geometric properties of the exact solution and has spectral accuracy in space.
Unfortunately, this algorithm cannot be used for all types of nonlinear
time-dependent Schr\"{o}dinger equations. Indeed, in the case of the
Gross-Pitaevskii equation, the algorithm is symmetric and, therefore,
time-reversible only because the ordinary differential equation that must be
solved when propagating the molecular state with the potential part of the
Hamiltonian leaves the nonlinear term invariant in
time.\cite{Bao_Markowich:2003} We show here that for nonlinear terms of more
general form, this algorithm becomes implicit. If this implicit nature is not
taken into account and the explicit version is used, the algorithm loses its
time reversibility and efficiency due to its low accuracy that is only of the
first order in the time step.

 An example of a situation, where a more general nonlinearity
appears, is provided by local control theory (LCT). Introduced by Kosloff
\textit{et al.},\cite{Kosloff_Tannor:1989} LCT is a widely used approach to
coherent control. In LCT, the pulse is computed on the fly, based on the
instantaneous molecular state, in order to increase (or decrease) an
expectation value of a specified operator. LCT has been successfully used to
control various processes such as energy and population
transfer,\cite{Kosloff_Tannor:1989, Kosloff_Tannor:1992,
Marquetand_Engel:2006a, Engel_Tannor:2009} dissociation and association
dynamics,\cite{Marquetand_Engel:2006b,
Marquetand_Engel:2007,Bomble_Desouter-Lecomte:2011,Vranckx_Desouter-Lecomte:2015}
direction of rotation in molecular rotors\cite{Yamaki_Fujimura:2005}, and
electron transfer.\cite{Vindel-Zandbergen_Sola:2016} Controlling quantum
systems using LCT changes the nature of the time-dependent Schr\"{o}dinger
equation. Because the time dependence of the pulse is determined exclusively
by the molecular state, the time-dependent Schr\"{o}dinger equation becomes
autonomous but nonlinear.

The nonlinear nature of LCT is often not acknowledged and the
standard explicit split-operator algorithm\cite{Feit_Steiger:1982} for linear
time-dependent Schr\"{o}dinger equations is used,\cite{Marquetand_Engel:2006b,
Marquetand_Engel:2007,Bomble_Desouter-Lecomte:2011,Vranckx_Desouter-Lecomte:2015,Vindel-Zandbergen_Sola:2016}
instead of its time-reversible, second-order, but implicit alternative. Most
previous studies used LCT for applications that required neither high
accuracy nor time reversibility, and therefore could rely on this approximate
explicit integrator, which, as we show below, in the context of LCT, indeed
has only first-order accuracy in the time step and is time-irreversible. Such
an algorithm, however, would be very inefficient for highly accurate
calculations, and could not be used at all if exact time reversibility were
important. Because this failure of the explicit splitting algorithm in LCT is
generic, while its success in the Gross-Pitaevskii equation is rather an
exception, it is desirable to develop efficient high-order geometric
integrators suitable for a general nonlinear time-dependent Schr\"{o}dinger
equation.

Recently, we presented high-order time-reversible geometric
integrators for the nonadiabatic quantum dynamics driven by the linear
time-dependent Schr\"{o}dinger equation with both
separable\cite{Roulet_Vanicek:2019} and nonseparable\cite{Choi_Vanicek:2019}
Hamiltonians. Here, we extend this work to the general nonlinear
Schr\"{o}dinger equation, in order to address the slow convergence and time
irreversibility of the explicit split-operator algorithm. 

The remainder of the study is organized as follows: In
Sec.~\ref{sec:theory_lct}, we define the nonlinear time-dependent
Schr\"{o}dinger equation, discuss its geometric properties, and explain how
LCT leads to a nonlinear Schr\"{o}dinger equation. In
Sec.~\ref{sec:geometic_integrators_NLTDSE}, after demonstrating the loss of
geometric properties by Euler methods, we describe how these geometric
properties are recovered and accuracy increased to an arbitrary even order by
symmetrically composing the implicit and explicit Euler methods. Then, we
describe a general procedure to perform the implicit propagation and derive
explicit expressions for the case of LCT. We also show the derivation of the
 approximate explicit split-operator algorithm for the nonlinear
Schr\"{o}dinger equation, explain how it loses time reversibility 
and briefly describe the dynamic Fourier method. Finally, in
Sec.~\ref{sec:numerical_examples} we numerically verify the convergence and
geometric properties of the integrators by controlling, using LCT, either the
population or energy transfer in a two-state two-dimensional model of
retinal.\cite{Hahn_Stock:2000}

\section{Nonlinear Schr\"{o}dinger equation}

\label{sec:theory_lct}

{The \emph{nonlinear time-dependent Schr\"{o}dinger equation} is
the differential equation
\begin{equation}
i\hbar\frac{d}{dt}|\psi_{t}\rangle=\hat{H}(\psi_{t})|\psi_{t}\rangle,
\label{eq:nl_TDSE}%
\end{equation}
describing the time evolution of the state $\psi_{t}$ driven by the nonlinear
Hamiltonian operator $\hat{H}(\psi_{t})$, which depends on the state of the
system. This dependence on $\psi_{t}$ is what distinguishes the equation from
the linear Schr\"{o}dinger equation. As the notation in Eq.~(\ref{eq:nl_TDSE})
suggests, we shall always assume that while the operator $\hat{H}:\psi
\mapsto\hat{H}(\psi)\psi$ is nonlinear, for each $\psi$ th{e operator $\hat
{H}(\psi):\phi\mapsto\hat{H}(\psi)\phi$ is linear. We will also assume that
$\hat{H}(\psi)$ has real expectation {values} $\langle\hat{H}(\psi)\rangle_{\phi
}:=\langle\phi|\hat{H}(\psi)\phi\rangle$ in any state $\phi$, which for a
linear operator implies that it is Hermitian, i.e., $\hat{H}(\psi)^{\dag}%
=\hat{H}(\psi)$, or, more precisely, that for every $\psi$, $\phi$, $\chi$,%
\begin{equation}
\langle\phi|\hat{H}(\psi)\chi\rangle=\langle\hat{H}(\psi)\phi|\chi\rangle.
\label{eq:hermiticity}%
\end{equation}
} }

{ A paradigm of a nonlinear Schr\"{o}dinger equation is the Gross-Pitaevskii
equation,\cite{Gross:1961, Pitaevskii:1961, Minguzzi_Vignolo:2004} in position
representation expressed as%
\[
i\hbar\partial_{t}\psi_{t}(q)=-\frac{\hbar^{2}}{2m}\nabla^{2}\psi
_{t}(q)+V(q)\psi_{t}(q)+C|\psi_{t}(q)|^{2}\psi_{t}(q),
\]
where the real coefficient $C$ is positive for a repulsive interaction and
negative for an attractive\ interaction. This equation has a cubic
nonlinearity and is useful, e.g., for approximate modeling of the dynamics of
a Bose-Einstein condensate.{\cite{Antoine_Besse:2013}} Many other examples are
provided by the Dirac-Frenkel variational principle,\cite{Dirac:1930,
book_Frenkel:1934} which approximates the exact solution of a linear
Schr\"{o}dinger equation with Hamiltonian $\hat{H}$ by an optimal solution of
a predefined, restricted form within a certain subset (called the
approximation manifold) of the Hilbert space. This optimal solution $\psi_{t}$
satisfies the equation%
\begin{equation}
\langle\delta\psi_{t}|(i\hbar d/dt-\hat{H})|\psi_{t}\rangle=0, \label{eq:DFVP}%
\end{equation}
where $\delta\psi_{t}$ is an arbitrary variation in the approximation
manifold. Equation (\ref{eq:DFVP})\ is equivalent to the nonlinear
Eq.~(\ref{eq:nl_TDSE}) with an effective state-dependent Hamiltonian%
\[
\hat{H}(\psi_{t})=\hat{P}(\psi_{t})\hat{H},
\]
where $\hat{P}(\psi_{t})$ is the projection operator on the tangent space to
the approximation manifold at the point $\psi_{t}$%
.\cite{book_Lubich:2008,Lasser_Lubich:2020} (Note that in the very special
case, where the projector does not depend on $\psi_{t}$, the resulting
Schr\"{o}dinger equation remains linear. This happens in the Galerkin method,
in which $\psi_{t}$ is expanded in a finite, time-independent basis and the
approximation manifold is a vector space.\cite{book_Lubich:2008})}

\subsection{Geometric properties of the exact evolution operator}

With initial condition $|\psi_{t_{0}}\rangle$ {and assuming that
$t\geq t_{0}$}, Eq.~(\ref{eq:nl_TDSE}) has the formal solution $|\psi
_{t}\rangle=\hat{U}(t,t_{0};\psi)|\psi_{t_{0}}\rangle$ with the exact
evolution operator given by
\begin{equation}
\hat{U}(t,t_{0};\psi):=\mathcal{T}\exp\left[  -\frac{i}{\hbar}\int_{t_{0}}%
^{t}dt^{\prime}\hat{H}(\psi_{t^{\prime}})\right]  ,
\label{eq:evol_oper_ex_LCT}%
\end{equation}
where the dependence of $\hat{U}$ on $\psi$ was added as an argument to
emphasize the nonlinear character of Eq.~(\ref{eq:nl_TDSE}).
{Expression (\ref{eq:evol_oper_ex_LCT}) is obtained by solving the
differential equation%
\[
i\hbar\frac{\partial}{\partial t}\hat{U}(t,t_{0};\psi)=\hat{H}(\psi_{t}%
)\hat{U}(t,t_{0};\psi)
\]
with initial condition $\hat{U}(t_{0},t_{0};\psi)=1$.{ The \emph{Hermitian
adjoint} of $\hat{U}(t,t_{0};\psi)$ is the operator%
\begin{equation}
\hat{U}(t,t_{0};\psi)^{\dag}=\mathcal{\bar{T}}\exp\left[  \frac{i}{\hbar}%
\int_{t_{0}}^{t}dt^{\prime}\hat{H}(\psi_{t^{\prime}})\right]  =\hat{U}%
(t,t_{0};\psi)^{-1}, \label{eq:U_adj}%
\end{equation}
where $\mathcal{\bar{T}}$ denotes the reverse time-ordering operator.} }

The nonlinearity of $\hat{H}$ leads to the loss of some geometric properties,
even if Eq.~(\ref{eq:nl_TDSE}) is solved exactly. Indeed, since the
Hamiltonian is nonlinear, the exact evolution operator is also nonlinear.

{ An operator $\hat{U}$ is said to \emph{preserve the inner product} (or to be
\emph{unitary}) if $\langle\hat{U}\psi|\hat{U}\phi\rangle=\langle\psi
|\phi\rangle$. The exact evolution operator does not preserve the inner
product because%
\begin{align}
\langle\psi_{t}|\phi_{t}\rangle &  =\langle\hat{U}(t,t_{0};\psi)\psi_{t_{0}%
}|\hat{U}(t,t_{0};\phi)\phi_{t_{0}}\rangle\nonumber\\
&  =\langle\psi_{t_{0}}|\hat{U}(t,t_{0};\psi)^{\dagger}\hat{U}(t,t_{0}%
;\phi)\phi_{t_{0}}\rangle\nonumber\\
&  =\langle\psi_{t_{0}}|\hat{U}(t,t_{0};\psi)^{-1}\hat{U}(t,t_{0};\phi
)\phi_{t_{0}}\rangle\nonumber\\
&  \neq\langle\psi_{t_{0}}|\phi_{t_{0}}\rangle
\end{align}
if $\psi_{t_{0}}\neq\phi_{t_{0}}${, where we used the property (\ref{eq:U_adj}) of the Hermitian adjoint of $\hat{U}$ to obtain the third line}. The exact nonlinear evolution operator is,
therefore, not \emph{symplectic} because it does not preserve the associated
symplectic two-form\cite{book_Lubich:2008}
\begin{equation}
\omega(\psi,\phi):=-2\hbar\operatorname{Im}\langle\psi|\phi\rangle.
\label{eq:two_form}%
\end{equation}

The nonlinear evolution does not conserve \emph{energy}%
\[
E_{t}:=\langle\hat{H}\left(  \psi_{t}\right)  \rangle_{\psi_{t}}=\langle
\psi_{t}|\hat{H}(\psi_{t})\psi_{t}\rangle,
\]
since%
\begin{align}
\frac{dE_{t}}{dt}  &  =\langle\dot{\psi}_{t}|\hat{H}(\psi_{t})|\psi_{t}%
\rangle+\langle\psi_{t}|\frac{d}{dt}\hat{H}(\psi_{t})|\psi_{t}\rangle
\nonumber\\
&  \qquad+\langle\psi_{t}|\hat{H}(\psi_{t})|\dot{\psi}_{t}\rangle\nonumber\\
&  =\langle d\hat{H}(\psi_{t})/dt\rangle_{\psi_{t}}\neq0,
\label{eq:exact_energy}%
\end{align}
where the first and third terms in the intermediate step cancel each other
because $\psi_{t}$ satisfies the nonlinear Schr\"{o}dinger equation
(\ref{eq:nl_TDSE}). Note, however, that in special cases, such as the
Gross-Pitaevskii equation, there exist modified energy functionals that are
conserved.{\cite{Antoine_Besse:2013}}

An operator $\hat{U}$ is said to \emph{conserve the norm} $\Vert\psi
\Vert:=\langle\psi|\psi\rangle^{1/2}$ if $\Vert\hat{U}\psi\Vert=\Vert\psi
\Vert$. The exact nonlinear evolution operator $\hat{U}\equiv\hat{U}%
(t,t_{0};\psi)$ conserves the norm because%
\begin{align}
\Vert\psi_{t}\Vert^{2}  &  =\Vert\hat{U}\psi_{t_{0}}\Vert^{2}=\langle\hat
{U}\psi_{t_{0}}|\hat{U}\psi_{t_{0}}\rangle\nonumber\\
&  =\langle\psi_{t_{0}}|\hat{U}^{\dagger}\hat{U}\psi_{t_{0}}\rangle
=\langle\psi_{t_{0}}|\psi_{t_{0}}\rangle=\Vert\psi_{t_{0}}\Vert^{2},
\label{eq:exact_norm}%
\end{align}
where we used relation (\ref{eq:U_adj}).

{ In the theory of dynamical systems, an \emph{adjoint} $\hat{U}(t,t_{0};\psi)^{\ast}$} of $\hat{U}(t,t_{0};\psi)$ is defined as the inverse of the
evolution operator taken {[in contrast to the Hermitian adjoint (\ref{eq:U_adj})]} with a reversed time flow:
\begin{equation}
\hat{U}(t,t_{0};\psi)^{\ast}:=\hat{U}(t_{0},t;\psi)^{-1}.
\end{equation}
An operator $\hat{U}(t,t_{0};\psi)$ is called \emph{symmetric} if it is equal
to its adjoint, i.e., if $\hat{U}(t,t_{0};\psi)=\hat{U}(t,t_{0};\psi)^{\ast}$.
A symmetric operator is also \emph{time-reversible} because propagating a
molecular state $\psi_{t_{0}}$ forward to time $t$ and then backward to time
$t_{0}$ yields $\psi_{t_{0}}$ again, i.e.,
\begin{align}
&  \hat{U}(t_{0},t;\psi)\hat{U}(t,t_{0};\psi)\psi_{t_{0}}\nonumber\\
&  =\hat{U}(t_{0},t;\psi)\hat{U}(t,t_{0};\psi)^{\ast}\psi_{t_{0}}\nonumber\\
&  =\hat{U}(t_{0},t;\psi)\hat{U}(t_{0},t;\psi)^{-1}\psi_{t_{0}}=\psi_{t_{0}}.
\label{eq:forward_backward}%
\end{align}
For the operator (\ref{eq:evol_oper_ex_LCT}), the reverse evolution operator
is given by the anti-time-ordered exponential
\begin{equation}
\hat{U}(t_{0},t;\psi):=\bar{\mathcal{T}}\exp\left[  -\frac{i}{\hbar}\int%
_{t}^{t_{0}}dt^{\prime}\hat{H}(\psi_{t^{\prime}})\right]
\end{equation}
and, therefore, the adjoint is
\begin{align}
\hat{U}(t,t_{0};\psi)^{\ast}  &  =\left\{  \bar{\mathcal{T}}\exp\left[
-\frac{i}{\hbar}\int_{t}^{t_{0}}dt^{\prime}\hat{H}(\psi_{t^{\prime}})\right]
\right\}  ^{-1}\nonumber\\
&  =\mathcal{T}\exp\left[  \frac{i}{\hbar}\int_{t}^{t_{0}}dt^{\prime}\hat
{H}(\psi_{t^{\prime}})\right] \nonumber\\
&  =\mathcal{T}\exp\left[  -\frac{i}{\hbar}\int_{t_{0}}^{t}dt^{\prime}\hat
{H}(\psi_{t^{\prime}})\right]  =\hat{U}(t,t_{0};\psi).
\end{align}
This shows that the exact evolution operator (\ref{eq:evol_oper_ex_LCT}) is
symmetric and time-reversible.

Finally, a time evolution is said to be \emph{stable} if for every
$\epsilon>0$ there is a $\delta>0$ such that the distance between two states
satisfies the condition
\begin{equation}
\Vert\psi_{t_{0}}-\phi_{t_{0}}\Vert<\delta\implies\Vert\psi_{t}-\phi_{t}%
\Vert<\epsilon\text{ for all }t>t_{0}.
\end{equation}
Although the exact evolution conserves the norm, because it does not conserve
the inner product, we cannot, in general, say anything about preserving the
distance:
\begin{align}
\Vert\psi_{t}-\phi_{t}\Vert^{2}  &  =\Vert\psi_{t}\Vert^{2}+\Vert\phi_{t}%
\Vert^{2}-2\operatorname{Re}\langle\psi_{t}|\phi_{t}\rangle\nonumber\\
&  =\Vert\psi_{t_{0}}\Vert^{2}+\Vert\phi_{t_{0}}\Vert^{2}-2\operatorname{Re}%
\langle\psi_{t}|\phi_{t}\rangle\nonumber\\
&  =\Vert\psi_{t_{0}}-\phi_{t_{0}}\Vert^{2}+2\operatorname{Re}\left(
\langle\psi_{t_{0}}|\phi_{t_{0}}\rangle-\langle\psi_{t}|\phi_{t}\rangle\right)
\nonumber\\
&  \neq\Vert\psi_{t_{0}}-\phi_{t_{0}}\Vert^{2}. \label{eq:exact_stability}%
\end{align}
Moreover, since the sign of the real part of the difference of the inner
products can be arbitrary, we cannot deduce anything about stability.}

\subsection{Nonlinear character of local control theory}

\label{subsec:lct}

{We now show that the local coherent control of the time evolution
of a quantum system with an electric field provides another example of a
nonlinear Schr\"{o}dinger equation. The} quantum state $|\psi_{t}\rangle$ of a
system interacting with an external time-dependent electric field $\vec{E}(t)$
evolves according to the linear time-dependent Schr\"{o}dinger equation
\begin{equation}
i\hbar\frac{d}{dt}|\psi_{t}\rangle=\hat{H}(t)|\psi_{t}\rangle\label{eq:TDSE}%
\end{equation}
with a time-dependent Hamiltonian
\begin{equation}
\hat{H}(t):=\hat{H}_{0}+\hat{V}_{\text{int}}(t), \label{eq:TD-ham}%
\end{equation}
equal to the sum of the Hamiltonian $\hat{H}_{0}$ of the system in the absence
of the field and the interaction potential $\hat{V}_{\text{int}}(t)$.
{Within the electric-dipole approximation}%
,\cite{Schatz_Ratner_book:2002} the interaction potential is
\begin{equation}
\hat{V}_{\text{int}}(t):=-\vec{\mu}(\hat{q})\cdot\vec{E}(t),
\label{eq:int_pot}%
\end{equation}
where the vector function $\vec{\mu}(\hat{q})$ { of the position
operator $\hat{q}$} denotes the electric-dipole operator of the system. Direct
integration of Eq.~(\ref{eq:TDSE}) with initial condition $|\psi_{t_{0}%
}\rangle$ leads to the formal solution $|\psi_{t}\rangle=\hat{U}(t,t_{0}%
)|\psi_{t_{0}}\rangle$ with the exact evolution operator given by the
time-ordered exponential
\begin{equation}
\hat{U}(t,t_{0}):=\mathcal{T}\exp\left[  -\frac{i}{\hbar}\int_{t_{0}}%
^{t}dt^{\prime}\hat{H}(t^{\prime})\right]  . \label{eq:evol_oper_ex}%
\end{equation}

This exact evolution operator has many important geometric properties: it is
linear, unitary, symmetric, time-reversible, and
stable.\cite{book_Halmos:1942, book_Leimkuhler_Reich:2004, book_Lubich:2008,
book_Hairer_Wanner:2006} Because it is unitary, the evolution operator
conserves the norm as well as the inner product and symplectic
structure.\cite{book_Lubich:2008} However, since the Hamiltonian is
time-dependent, the Schr\"{o}dinger equation~(\ref{eq:TDSE}) is a
nonautonomous differential equation,\cite{book_Hairer_Wanner:2006} and as a
consequence, does not conserve energy. For a more detailed presentation and
discussion of the above properties, we refer the reader to Ref.~\onlinecite{Choi_Vanicek:2019}.

Contrary to Eq.~(\ref{eq:int_pot}), the electric field used in LCT, called
control field and denoted by $\vec{E}_{\text{LCT}}(t)$, is not known
explicitly as a function of time. Instead, it is chosen \textquotedblleft on
the fly\textquotedblright\ according to the current state $|\psi_{t}\rangle$
of the system, in order to increase or decrease the expectation value
$\langle\hat{O}\rangle_{{\psi}_{t}}:=\langle\psi_{t}|\hat{O}|\psi_{t}\rangle$
of a particular operator $\hat{O}$ in the state $|\psi_{t}\rangle$. More
precisely, the control field is computed so that the time derivative of the
expectation value,
\begin{align}
\frac{d\langle\hat{O}\rangle_{\psi_{t}}}{dt}  &  =\frac{i}{\hbar}%
\langle\lbrack\hat{H}(t),\hat{O}]\rangle_{\psi_{t}}\nonumber\\
&  =\frac{i}{\hbar}\left\{  \langle\lbrack\hat{H}_{0},\hat{O}]\rangle
_{\psi_{t}}-\vec{E}_{\text{LCT}}(t)\cdot\langle\lbrack\hat{\vec{\mu}},\hat
{O}]\rangle_{\psi_{t}}\right\}  , \label{eq:O_derivative}%
\end{align}
remains positive or negative at all times. If the operator $\hat{O}$ commutes
with the system's Hamiltonian $\hat{H}_{0}$, this goal is achieved by using
the field
\begin{equation}
\vec{E}_{\text{LCT}}(t)\equiv\vec{E}_{\text{LCT}}(\psi_{t}):=\pm\lambda
i\langle\lbrack\hat{\vec{\mu}},\hat{O}]\rangle_{\psi_{t}}^{\ast}=\mp\lambda
i\langle\lbrack\hat{\vec{\mu}},\hat{O}]\rangle_{\psi_{t}},
\label{eq:LCT_field}%
\end{equation}
where $\lambda>0$ is a parameter which scales the intensity of the control
field and the sign in Eq.~(\ref{eq:LCT_field}) is chosen according to whether
one wants to increase or decrease $\langle\hat{O}\rangle_{\psi_{t}}$. This
claim is proven by inserting the definition~(\ref{eq:LCT_field}) of $\vec
{E}_{\text{LCT}}(t)$ into Eq.~(\ref{eq:O_derivative}), which yields
\begin{equation}
\frac{d\langle\hat{O}\rangle_{\psi_{t}}}{dt}=\frac{i}{\hbar}\langle\lbrack
\hat{H}_{0},\hat{O}]\rangle_{\psi_{t}}\pm\frac{\lambda}{\hbar}\Vert
\langle\lbrack\hat{\vec{\mu}},\hat{O}]\rangle_{\psi_{t}}\Vert^{2}
\label{eq:O_derivative_simplified}%
\end{equation}
for the derivative of the expectation value. This equation confirms that a
strictly increasing or strictly decreasing evolution of $\langle\hat{O}%
\rangle_{\psi_{t}}$ is guaranteed only if $[\hat{H}_{0},\hat{O}]=0$, largely
reducing the choice of operators $\hat{O}$ whose expectation values we can
control monotonically.

The left-hand side of Eq.~(\ref{eq:LCT_field}) suggests that the control field
can be either viewed as a function of time or a function of the molecular
state [i.e., $\vec{E}_{\text{LCT}}(t)\equiv\vec{E}_{\text{LCT}}(\psi_{t})$].
More precisely, the control field does not depend on time explicitly but only
implicitly through the dependence on $\psi_{t}$. Therefore, the time-dependent
Schr\"{o}dinger equation changes from a nonautonomous linear to an autonomous
nonlinear differential equation.\cite{book_Hairer_Wanner:2006} By
acknowledging this nonlinear character, the interaction potential from Eq.~(\ref{eq:int_pot}) becomes
\begin{equation}
\hat{V}_{\text{LCT}}(\psi_{t}):=-\hat{\vec{\mu}}\cdot\vec{E}_{\text{LCT}}%
(\psi_{t}) \label{eq:nl_int_pot}%
\end{equation}
and Eq.~(\ref{eq:TDSE}) becomes an example of a nonlinear time-dependent
Schr\"{o}dinger equation (\ref{eq:nl_TDSE}) with Hamiltonian operator $\hat
{H}(\psi):=\hat{H}_{0}+\hat{V}_{\text{LCT}}(\psi)$.

\section{Geometric integrators for the nonlinear Schr\"{o}dinger equation}

\label{sec:geometic_integrators_NLTDSE}

Numerical propagation methods for solving the nonlinear equation
(\ref{eq:nl_TDSE}) obtain the state at time $t+\Delta t$ from the state at
time $t$ by using the relation
\begin{equation}
|\psi_{t+\Delta t}\rangle=\hat{U}_{\text{appr}}(t+\Delta t,t;\psi)|\psi
_{t}\rangle,
\end{equation}
where $\Delta t$ denotes the numerical time step and $\hat{U}_{\text{appr}%
}(t+\Delta t,t;\psi)$ is an approximate nonlinear evolution operator depending
on $\psi$. { By construction, all propagation methods converge to
the exact solution in the limit $\Delta t\rightarrow0$. As the exact operator,
these approximate evolution operators $\hat{U}_{\text{appr}}(t+\Delta
t,t;\psi)$ are, therefore, nonlinear and conserve neither the inner product
nor the symplectic form. Moreover, nothing can be said about their stability
in general. However, some integrators may lose even the remaining geometric
properties of the exact evolution: norm conservation, symmetry, and time
reversibility. In this section, we simply state the properties that are lost
by different methods; detailed proofs are provided in Appendix
\ref{sec:proof_geometric_prop}. }

\subsection{Loss of geometric properties by Euler methods}

\label{subsec:lct_euler_methods}

The simplest methods, applicable to both separable and nonseparable and both
linear and nonlinear Hamiltonian operators, are the \emph{explicit} and
\emph{implicit} Euler\cite{book_Lubich:2008, book_Leimkuhler_Reich:2004}
methods which approximate the exact evolution operator, respectively, as
\begin{align}
\hat{U}_{\text{expl}}(t+\Delta t,t;\psi_{t})  &  :=1-i\hat{H}(\psi_{t})\Delta
t/\hbar,\label{eq:expl}\\
\hat{U}_{\text{impl}}(t+\Delta t,t;\psi_{t+\Delta t})  &  :=[1+i\hat{H}%
(\psi_{t+\Delta t})\Delta t/\hbar]^{-1}. \label{eq:impl}%
\end{align}
Both methods are only first-order in the time step and, therefore, very
inefficient. { Moreover, both Euler methods lose the norm
conservation, symmetry, and time reversibility of the exact evolution
operator.}

\subsection{Recovery of geometric properties and increasing accuracy by
composition}

\label{subsec:lct_geom_methods}

Composing the implicit and explicit Euler methods yields, {depending on the
order of composition,} either the \emph{implicit midpoint} method
\begin{multline}
\hat{U}_{\text{mid}}(t+\Delta t,t;\psi_{t+\Delta t/2})\label{eq:mid}\\
:=\hat{U}_{\text{expl}}(t+\Delta t,t+\Delta t/2;\psi_{t+\Delta t/2})\\
\times\hat{U}_{\text{impl}}(t+\Delta t/2,t;\psi_{t+\Delta t/2})
\end{multline}
{or the \emph{trapezoidal rule} (or \emph{Crank--Nicolson} method)
}
\begin{multline}
\hat{U}_{\text{trap}}(t+\Delta t,t;\psi)\label{eq:trap}\\
:=\hat{U}_{\text{impl}}(t+\Delta t,t+\Delta t/2;\psi_{t+\Delta t})\\
\times\hat{U}_{\text{expl}}(t+\Delta t/2,t;\psi_{t}).
\end{multline}
{Both methods are second-order, symmetric, and time-reversible
regardless of the size of the time
step.\cite{book_Hairer_Wanner:2006,Choi_Vanicek:2019} Although the trapezoidal
rule conserves the norm of a state evolved with a time-independent linear
Hamiltonian,\cite{Choi_Vanicek:2019} it loses this property when the
Hamiltonian is time-dependent or nonlinear (which results, as we have seen, in
an implicit time dependence). In contrast, the implicit midpoint method
remains norm-conserving in all cases.}

Because they are symmetric, both methods can be further composed using
symmetric composition schemes\cite{book_Lubich:2008, book_Hairer_Wanner:2006,
Yoshida:1990, Suzuki:1990, Kahan_Li:1997,
Sofroniou_Spaletta:2005,Choi_Vanicek:2019} in order to obtain integrators of
arbitrary even order of accuracy in the time step. Indeed, every symmetric
method $\hat{U}_{p}$ of an even order $p$ generates a method $\hat{U}_{p+2}$
of order $p+2$ if it is symmetrically composed as
\begin{multline}
\hat{U}_{p+2}(t+\Delta t,t;\psi):=\hat{U}_{p}(t+\gamma_{M}\Delta
t,t+\gamma_{M-1}\Delta t;\psi)\nonumber\\
\quad\cdots\hat{U}_{p}(t+\gamma_{1}\Delta t,t;\psi),
\end{multline}
where $M$ is the number of composition steps, $\gamma_{1},\dots,\gamma_{M}$
are real composition coefficients which satisfy the relations $\sum_{n=1}%
^{M}\gamma_{n}=1$, $\gamma_{M+1-n}=\gamma_{n},$ and a more-involved third
condition\cite{book_Hairer_Wanner:2006} guaranteeing the increase in the order
of accuracy.

{The simplest composition methods that are used here are the
triple-jump~\cite{Yoshida:1990} and Suzuki's fractal\cite{Suzuki:1990} with
$M=3$ and $M=5$, respectively. Both are recursive and able to generate
integrators of arbitrary even orders of accuracy. Sixth, eighth, and
tenth-order integrators can also be obtained with nonrecursive composition
methods\cite{Kahan_Li:1997, Sofroniou_Spaletta:2005} which require fewer
composition steps and also minimize their magnitude. These composition methods
are therefore more efficient than the triple-jump and Suzuki's fractal and
will be referred to as \textquotedblleft optimal\textquotedblright%
\ compositions. For more details on symmetric compositions, see %
Ref.~\onlinecite{Choi_Vanicek:2019}, where they were applied to a linear
Schr\"{o}dinger equation.}

\subsection{Solving the implicit step in a general nonlinear Schr\"{o}dinger
equation}

\label{subsec:implicit_step}

The implicit Euler method requires implicit propagation because its integrator
[see Eq.~(\ref{eq:impl})] depends on the result of the propagation, i.e.,
$\psi_{t+\Delta t}$. In the implicit Euler method, $\psi_{t+\Delta t}$ is
obtained by solving the nonlinear system
\begin{equation}
\hat{U}_{\text{impl}}(t+\Delta t,t;\psi_{t+\Delta t})^{-1}|\psi_{t+\Delta
t}\rangle=|\psi_{t}\rangle,
\end{equation}
which can be written as $f(\psi_{t+\Delta t})=0$ with the nonlinear
functional
\begin{align}
f(\psi):  &  =\hat{U}_{\text{impl}}(\psi)^{-1}\psi-\psi_{t}\nonumber\\
&  =[\hat{1}+i\hat{H}(\psi)\Delta t/\hbar]\psi-\psi_{t}. \label{eq:nl_system}%
\end{align}
A nonlinear system $f(\psi)=0$ can be solved with the iterative Newton-Raphson
method, which computes, until convergence is obtained, the solution
$\psi^{(k+1)}$ at iteration $k+1$ from $\psi^{(k)}$ using the relation%
\begin{equation}
\psi^{(k+1)}=\psi^{(k)}-\hat{J}(\psi^{(k)})^{-1}f(\psi^{(k)}),
\label{eq:newton-raphson}%
\end{equation}
where $\hat{J}:=\frac{\delta}{\delta\psi}f(\psi)$ is the Jacobian of the
nonlinear functional $f(\psi)$.

If the initial guess $\psi^{(0)}$ is close enough to the exact solution of the
implicit propagation, the Newton-Raphson iteration~(\ref{eq:newton-raphson})
is a contraction mapping and by the fixed-point theorem is guaranteed to
converge. We use as the initial guess the result of propagating $\psi_{t}$
with the explicit Euler method [Eq.~(\ref{eq:expl})]. Note that this initial
guess is sufficiently close to the implicit solution only if the time step is
small. If the time step is too large, the difference between the explicit and
implicit propagations becomes too large for the algorithm to converge and no
solution can be obtained.

Equation (\ref{eq:newton-raphson}) requires computing the inverse of the
Jacobian which is an expensive task. It is preferable to avoid this inversion
by computing each iteration as
\begin{equation}
\psi^{(k+1)}=\psi^{(k)}+\delta\psi^{(k)} \label{eq:iteration}%
\end{equation}
where $\delta\psi^{(k)}$ solves the linear system
\begin{equation}
\hat{J}(\psi^{(k)})\delta\psi^{(k)}=-f(\psi^{(k)}). \label{eq:linear_system}%
\end{equation}
We solve this linear system by the generalized minimal residual
method~\cite{book_Press_Flannery:1992, Saad_Schultz:1986, book_Saad:2003}, an
iterative method based on the Arnoldi process\cite{Arnoldi:1951,Saad:1980}
{(see Sec.~I of the supplementary material for a
detailed presentation of this algorithm).}

The procedure presented for solving the implicit propagation is applicable to
any nonlinear system whose Jacobian is known analytically. Therefore, the
integrators proposed in Secs.~\ref{subsec:lct_euler_methods} and
\ref{subsec:lct_geom_methods} can be employed for solving any nonlinear
time-dependent Schr\"{o}dinger equation of the form of Eq.~(\ref{eq:nl_TDSE}),
i.e., with a Hamiltonian $\hat{H}(\psi_{t})$ depending on the state of the system.

{To sum up, each implicit propagation step, given by the evolution
operator (\ref{eq:impl}), is performed as follows:}

\begin{enumerate}
\item {Compute the initial guess $\psi^{(0)}$ using the explicit
Euler method [see Eq.~(\ref{eq:expl})]. Choose an error threshold
$\varepsilon$ and a maximum iteration number $m$.}

\item {For $k=0,1,\dots,m-1$, Do:} {\setlength\itemindent{2em}}

\item {Compute $\delta\psi^{(k)}$ by solving the linear system
shown in Eq.~(\ref{eq:linear_system}).\label{item:solve_linear}}
{\setlength\itemindent{2em}}

\item {Compute a new approximate solution $\psi^{(k+1)}$ using
Eq.~(\ref{eq:iteration}).} {\setlength\itemindent{2em} }

\item {If $\Vert f(\psi^{(k+1)})\Vert\leq\varepsilon$, take
$\psi^{(k+1)}$ as the solution of the implicit step.}

\item {End Do. The algorithm fails when $k=m$ and no approximate
solution has been found.}
\end{enumerate}

\subsection{Solving the implicit step in LCT}

In the case of LCT,
\[
\hat{U}_{\text{LCT,impl}}(\psi)^{-1}=\hat{1}+i\Delta t[\hat{H}_{0}+\hat
{V}_{\text{LCT}}(\psi)]/\hbar,
\]
and the Jacobian of the nonlinear functional (\ref{eq:nl_system}) is%
\begin{align}
\hat{J}(\psi)  &  =\frac{\delta}{\delta\psi}\left[  \hat{U}_{\text{LCT,impl}%
}(\psi)^{-1}\right]  \psi+\hat{U}_{\text{LCT,impl}}(\psi)^{-1}\hat
{1}\nonumber\\
&  =\frac{i}{\hbar}\Delta t\frac{\delta}{\delta\psi}\left[  \hat
{V}_{\text{LCT}}(\psi)\right]  \psi+\hat{U}_{\text{LCT,impl}}(\psi
)^{-1}\nonumber\\
&  =\frac{i}{\hbar}\Delta t\hat{V}_{\text{LCT}}(\psi)+\hat{1}+\frac{i}{\hbar
}\Delta t\left[  \hat{H}_{0}+\hat{V}_{\text{LCT}}(\psi)\right] \nonumber\\
&  =\hat{1}+i\Delta t[\hat{H}_{0}+2\hat{V}_{\text{LCT}}(\psi)]/\hbar.
\label{eq:Jacobian_impl}%
\end{align}
To obtain the third row of Eq.~(\ref{eq:Jacobian_impl}), we used $\frac
{\delta}{\delta\psi}[\hat{V}_{\text{LCT}}(\psi)]\psi=\hat{V}_{\text{LCT}}%
(\psi)$, where the generalized complex derivative\cite{Petersen_Pedersen:2012}
of the interaction potential is given by {the bra vector}
\begin{equation}
\frac{\delta}{\delta\psi}\hat{V}_{\text{LCT}}(\psi)=-\hat{\vec{\mu}}\cdot
\frac{\delta}{\delta\psi}\vec{E}_{\text{LCT}}(\psi)=\mp\lambda i\hat{\vec{\mu
}}\cdot\langle\psi|[\hat{\vec{\mu}},\hat{O}].
\end{equation}

\subsection{Approximate application of the explicit split-operator algorithm
to the nonlinear Schr\"{o}dinger equation}

\label{subsec:naive_so}

The algorithms that we described above apply to Hamiltonians that are not only
nonlinear but also nonseparable, i.e., to Hamiltonians $\hat{H}$ which cannot
be written as a sum $\hat{H}=T(\hat{p})+V(\hat{q})$ of a momentum-dependent
kinetic term and position-dependent potential term. If the time-dependent
Schr\"{o}dinger equation is linear and its Hamiltonian is separable, the
midpoint method remains implicit, but the split-operator algorithms and their
compositions yield explicit high-order integrators satisfying most geometric
properties (except for the conservation of energy). In the case of LCT, if
$\hat{H}_{0}$ is separable, so is the total Hamiltonian, which can be written
as $\hat{H}(\psi)=\hat{T}+\hat{V}_{\text{tot}}(\psi)$, where $\hat
{V}_{\text{tot}}(\psi):=\hat{V}_{0}+\hat{V}_{\text{LCT}}(\psi)$ is the sum of
the system's and interaction potential energy operators. It is, therefore,
tempting to use the split-operator algorithm, with the hope of obtaining an
efficient explicit integrator.

More generally, let us assume that the Hamiltonian operator in the general
nonlinear Schr\"{o}dinger equation (\ref{eq:nl_TDSE}) can be separated as%
\[
\hat{H}(\psi)=T(\hat{p})+V_{\text{tot}}(\hat{q},\psi).
\]
The approximate evolution operator is given by
\begin{equation}
\hat{U}_{\text{TV}}(t+\Delta t;t,\psi_{t}):=e^{-i\hat{T}\Delta t/\hbar
}e^{-i\hat{V}_{\text{tot}}(\psi_{t})\Delta t/\hbar} \label{eq:TV}%
\end{equation}
in the \emph{TV} split-operator algorithm and by
\begin{equation}
\hat{U}_{\text{VT}}(t+\Delta t;t,\psi_{t+\Delta t}):=e^{-i\hat{V}_{\text{tot}%
}(\psi_{t+\Delta t})\Delta t/\hbar}e^{-i\hat{T}\Delta t/\hbar} \label{eq:VT}%
\end{equation}
in the \emph{VT} split-operator algorithm. These integrators are
norm-conserving but only first-order and time-irreversible. From their
definitions (\ref{eq:TV}) and (\ref{eq:VT}), it follows immediately that the
TV and VT algorithms are adjoints\cite{Roulet_Vanicek:2019} of each other and
require, respectively, explicit and implicit propagations. In analogy to the
implicit midpoint algorithm from Sec.~\ref{subsec:lct_geom_methods}, a
second-order method is obtained by composing the two adjoint methods to obtain
the \emph{TVT }split-operator algorithm
\begin{multline}
\hat{U}_{\text{TVT}}(t+\Delta t;t,\psi_{t+\Delta t/2})\label{eq:TVT}\\
:=\hat{U}_{\text{TV}}(t+\Delta t;t+\Delta t/2,\psi_{t+\Delta t/2})\\
\times\hat{U}_{\text{VT}}(t+\Delta t/2;t,\psi_{t+\Delta t/2}),
\end{multline}
or the \emph{VTV} split-operator algorithm if the order of composition is
reversed. Both TVT and VTV algorithms are norm-conserving, symmetric, and
time-reversible. However, these geometric properties are only acquired if the
implicit part, i.e., the propagation with the VT algorithm~(\ref{eq:TVT}) is
performed exactly. This requires solving a nonlinear system, which can be
performed using the Newton-Raphson method, as described in
Sec.~\ref{subsec:implicit_step}. This, however, implies abandoning the
explicit nature of the split-operator algorithm, which is one of its main
advantages over implicit methods for solving linear Schr\"{o}dinger equations.

{The nonlinearity of Eq.~(\ref{eq:nl_TDSE}) is often not
acknowledged. As a consequence, the implicit character of Eqs.~(\ref{eq:VT}%
)-(\ref{eq:TVT}) is not taken into account and explicit alternatives of these
algorithms are used. For example, instead of using $\psi_{t+\Delta t}$ in the
VT algorithm (i.e., performing the implicit propagation exactly), the state
$\psi_{t,\hat{T}\Delta t}:=e^{-i\hat{T}\Delta t/\hbar}\psi_{t}$  obtained
after the kinetic propagation is often used to perform the potential
propagation.After composition with the TV algorithm, it yields the \emph{approximate explicit} TVT split-operator algorithm}
\begin{multline}
\hat{U}_{\text{expl TVT}}(t+\Delta t;t,{\psi_{t,\hat{T}\Delta t/2}%
})\label{eq:TVT_naive}\\
:=\hat{U}_{\text{TV}}(t+\Delta t;t+\Delta t/2,{\psi_{t,\hat{T}\Delta t/2}})\\
\times\hat{U}_{\text{VT}}(t+\Delta t/2;t,{\psi_{t,\hat{T}\Delta t/2}}).
\end{multline}
{This approximate explicit integrator can be used to successfully
perform LCT but, because it depends on $\psi_{t,\hat{T}\Delta t/2}$ instead of
$\psi_{t+\Delta t/2}$, it is not time-reversible and achieves only first-order
accuracy. Like the approximate explicit TVT algorithm, any other explicit
algorithm cannot perform the implicit step of the $\hat{U}_{VT}$ algorithm
exactly and, therefore, cannot be time-reversible. Yet, the
approximate explicit algorithm does conserve norm (see Appendix
\ref{sec:proof_geometric_prop} for proofs of the geometric properties).}

\subsection{{Dynamic Fourier method}}

To propagate the molecular state using the algorithms presented in
Sec.~\ref{sec:geometic_integrators_NLTDSE}, an efficient way of evaluating the
action of an operator on the molecular state $\psi_{t}$ is needed. For this,
we use the dynamic Fourier method,\cite{Feit_Steiger:1982,
book_Tannor:2007,Kosloff_Kosloff:1983, Kosloff_Kosloff:1983b} which is
applicable to operators that are sums of products of operators of the form
$f(\hat{x})$, where $\hat{x}$ denotes either the momentum operator $\hat{p}$
or position operator $\hat{q}$. The computation of $f(\hat{x})\psi_{t}$ is
then performed easily, by pointwise multiplication, in the $x$-representation,
in which $\hat{x}$ is diagonal. Whenever required, the representation of
$\psi_{t}$ is changed by Fourier transform. In the numerical examples below,
we used the Fastest Fourier Transform in the West 3 (FFTW3)
library~\cite{Frigo_Johnson:2005} to perform all of the Fourier transforms.
Unlike finite difference methods, the Fourier method shows exponential
convergence with respect to the number of grid points (see Figs.~S1 and S2 of
the supplementary material).

\section{Numerical examples}

\label{sec:numerical_examples}We tested the general integrators for the
nonlinear Schr\"{o}dinger equation, presented in
Sec.~\ref{sec:geometic_integrators_NLTDSE}, by using them for the local
control of a two-dimensional two-state diabatic model of retinal taken from
Ref.~\onlinecite{Hahn_Stock:2000}. The model describes the cis-trans
photo-induced isomerization of retinal---an ultrafast reaction mediated by a
conical intersection and the first event occurring in the biological process
of vision. The two vibrational modes of the model are the reaction coordinate
$\theta$, an angle describing the torsional motion of the retinal molecule,
and a vibronically active coupling mode $q_{c}$. In the diabatic
representation, the Hamiltonian of the system in the absence of the field,
\begin{equation}
\hat{\mathbf{H}}_{0}=\hat{T}\mathbf{1}+%
\begin{pmatrix}
V_{11}(\hat{q}_{c},\hat{\theta}) & V_{12}(\hat{q}_{c})\\
V_{21}(\hat{q}_{c}) & V_{22}(\hat{q}_{c},\hat{\theta})
\end{pmatrix}
,
\end{equation}
is separable into a sum of the kinetic energy operator
\begin{equation}
\hat{T}=-\frac{1}{2}\omega\frac{\partial^{2}}{\partial q_{c}^{2}}-\frac{1}%
{2}m^{-1}\frac{\partial^{2}}{\partial\theta^{2}}%
\end{equation}
and potential energy operator with components%
\begin{align}
V_{11}(q_{c},\theta)  &  =\frac{1}{2}\omega q_{c}^{2}+\frac{1}{2}W_{1}\left[
1-\cos(\theta)\right]  ,\label{eq:pot_11}\\
V_{22}(q_{c},\theta)  &  =\frac{1}{2}\omega q_{c}^{2}+\chi_{2}q_{c}%
+E_{2}-\frac{1}{2}W_{2}\left[  1-\cos(\theta)\right]  ,\label{eq:pot_22}\\
V_{12}(q_{c})  &  =V_{21}(q_{c})=\xi q_{c}. \label{eq:couplings}%
\end{align}
Here (all parameters are in eV units), $\omega=0.19$ is the vibrational
frequency of the coupling mode, $m^{-1}=4.84\cdot10^{-4}$ is the inverse mass
of the reaction coordinate, $W_{1}=3.6$ and $W_{2}=1.09$ determine the depth
of the well in the reaction coordinate for the ground and excited electronic
states, respectively, $\chi_{2}=0.1$ is the gradient of the linear
perturbation in the excited electronic state, $E_{2}=2.48$ determines the
maximum of the excited electronic state in the reaction coordinate, and
$\xi=0.19$ is the gradient of the linear coupling between the two electronic
states. {The two diabatic potential energy surfaces
(\ref{eq:pot_11}) and (\ref{eq:pot_22}) are displayed in Fig.~S3 of the
supplementary material.}

{Note that, to distinguish nuclear, electronic, and molecular
operators, in this section} the \textbf{bold} face denotes electronic
operators expressed as $S\times S$ matrices in the basis of $S$ electronic
states and the hat $\hat{}$ denotes nuclear operators acting on the Hilbert
space of nuclear wavefunctions, i.e., square-integrable functions of $D$
continuous dimensions.

{ In the simulations, the reaction and coupling coordinates are
represented on regular grids consisting of 128 points between $\theta=-\pi/2$~a.u. and $\theta=\pi/2$~a.u.
and 64 points between $q_{c}=-9~$a.u. and $q_{c}=9~$a.u. Figure~S1 of the
supplementary material confirms that this grid is sufficient by showing
that the grid representation of the wavepacket is converged even at the final
time $t_{f}$. We assume the coordinate independence of the
electric dipole moment operator (Condon approximation) and, therefore, can
write $\hat{\vec{\bm{\mu}}}=\vec{\bm{\mu}}\hat{1}=\vec{\epsilon}%
\bm{\mu}\hat{1}$, where $\vec{\epsilon}$ is a constant unit vector in the
direction of $\vec{\bm{\mu}}$. In this case, the LCT electric field is aligned
with $\vec{\bm{\mu}}$ and we can write it as $\vec{E}_{\text{LCT}}%
=\vec{\epsilon}E_{\text{LCT}}$. As a consequence, we can drop the vector
symbol $\vec{}$ from $\hat{\vec{\bm{\mu}}}$ and $\vec{E}_{\text{LCT}}$ in
Eqs.~(\ref{eq:LCT_field}) and (\ref{eq:nl_int_pot}) and consider only the
analogous scalar equations satisfied by $\hat{\bm{\mu}}$ and $E_{\text{LCT}}$.
In addition, we assume the electric dipole moment operator to have unit
transition elements ($\hat{\mu}_{12}=\hat{\mu}_{21}=\hat{\mu}=1$~a.u.) and
zero diagonal elements ($\hat{\mu}_{11}=\hat{\mu}_{22}=0$).} The calculations
presented below aim to simulate the photo-excitation step of the
photo-isomerization of the retinal molecule. We therefore use as initial state
$\psi_{0}$ the ground vibrational state of the harmonic fit of the ground
potential energy surface [i.e., a two-dimensional Gaussian wavepacket with
$q_{0}=p_{0}=(0,0)$, and $\sigma_{0}=(0.128,1)$~a.u.] with initial populations
$P_{1}(0)=0.999$ and $P_{2}(0)=0.001$ of the ground and excited electronic
states, respectively. The tiny initial seed population of the excited state is
essential for the control because it ensures that Eq.~(\ref{eq:LCT_field})
does not stay zero at all times.

Two ways of populating the excited state based on LCT were investigated: the
former used as the target observable the population of the excited state
described by the projection operator onto the excited state (i.e.,
$\hat{\mathbf{O}}=\hat{\mathbf{P}}_{2}=\mathbf{P}_{2}\hat{1}=\mathbf{P}_{2}$),
while the latter employed as the target observable the molecular energy
described by the unperturbed molecular Hamiltonian (i.e., $\hat{\mathbf{O}%
}=\hat{\mathbf{H}}_{0}$). {To show that the monotonic evolution of
the target observable $\langle\mathbf{\hat{O}}\rangle_{\psi_{t}}$ is
guaranteed only if $[\mathbf{\hat{O}},\mathbf{\hat{H}}_{0}]=0$, we also
compare the results obtained from control calculations in the presence of
nonadiabatic couplings (where $[\hat{\mathbf{P}}_{2},\hat{\mathbf{H}}_{0}%
]\neq0$ and $[\hat{\mathbf{H}}_{0},\hat{\mathbf{H}}_{0}]=0$) and in the
absence of nonadiabatic couplings (where both target operators %
$\hat{\mathbf{P}}_{2}$ and {$\mathbf{\hat{H}}_{0}$ commute with $\mathbf{\hat
{H}}_{0}$).}} The control calculations were performed by solving the nonlinear
time-dependent Schr\"{o}dinger equation~(\ref{eq:nl_TDSE}) with the implicit
midpoint algorithm combined with the dynamic Fourier
method\cite{Feit_Steiger:1982, book_Tannor:2007, Kosloff_Kosloff:1983,
Kosloff_Kosloff:1983b} for a total time $t_{f}=256$~a.u. with a time step
$\Delta t=2^{-3}$~a.u. In addition, intensity parameters $\lambda
=1.430\times10^{-2}$ and $\lambda=1.534\times10^{-1}$ were used for the
control of excited-state population $P_{2}(t)=\langle\mathbf{P}_{2}%
\rangle_{\psi_{t}}$ and molecular energy $E_{0}(t)=\langle\mathbf{\hat{H}}%
_{0}\rangle_{\psi_{t}}$, respectively. These parameters were chosen so that
the electric fields of the obtained control pulses were similar during the
first period.

\begin{figure}
[htb]\centering\includegraphics[width=\linewidth]{
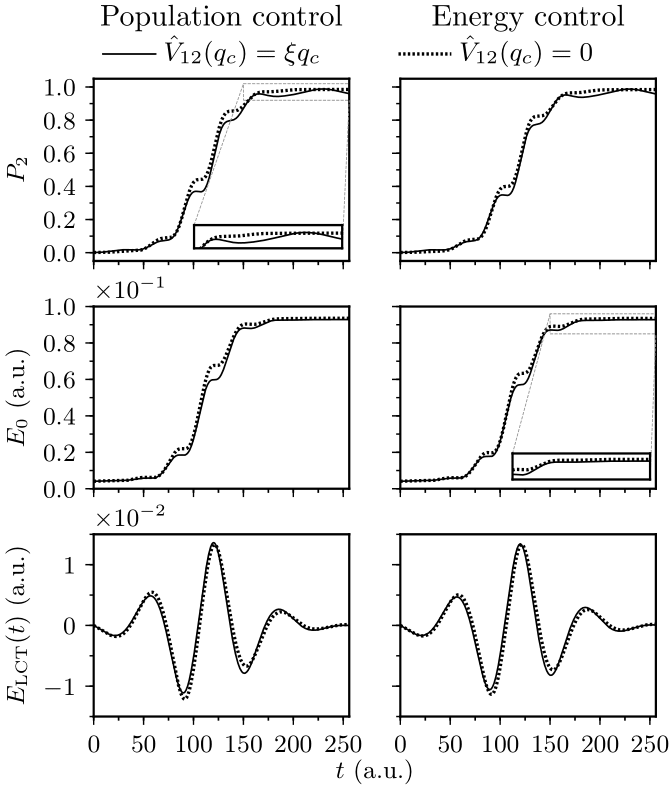}
\caption{Local control
calculations whose goal is increasing either the population
$P_{2}:=\langle \mathbf{P}_{2}\rangle_{\psi_{t}}$  of the excited state
(left panels, $\lambda=1.430 \times 10^{-2}$) or the molecular energy
$E_{0}(t):=\langle\mathbf{\hat{H}}_{0}\rangle_{\psi_{t}}$ (right panels,
$\lambda = 1.534\times 10^{-1}$). As expected, the local control theory
applied to these closely related objectives yields very similar results.
{Top: Excited state population. Middle: Molecular energy. Bottom: Pulse
obtained by local control theory. }} \label{fig:retinal_control}
\end{figure}

{Figure~\ref{fig:retinal_control} shows the excited-state
population, molecular energy, and obtained control pulse for the control of
either the excited-state population (left panels) or molecular energy (right
panels).} In the figure, the results obtained in the presence and in the
absence of nonadiabatic couplings are also compared for each target. The
population and energy control schemes result in similar population dynamics
and in both schemes, the population of the excited state reaches $0.99$ at
time $t_{f}$. { The carrier frequencies of the obtained control
pulses are, as expected, similar and correspond to the electronic transition
between the two electronic states of the model.} As predicted, when
controlling the excited-state population $\langle\mathbf{P}_{2}\rangle
_{\psi_{t}}$ in the presence of nonadiabatic couplings { given by
Eq.~(\ref{eq:couplings})}, the evolution of the population is not monotonic
{(see the solid line in the inset of the top left panel of
Fig.~\ref{fig:retinal_control})} because the control operator does not commute
with the molecular Hamiltonian. { Compare this with the monotonic
increase of the population when the nonadiabatic couplings are zero [dotted
line in the same inset; $\hat{V}_{12}(q_{c})=\hat{V}_{21}(q_{c})=0$] and the
target operator commutes with the molecular Hamiltonian. In contrast, when
controlling the molecular energy $\langle\hat{\mathbf{H}}_{0}\rangle_{\psi
_{t}}$, its time evolution is always monotonic because the molecular
Hamiltonian commutes with itself, whether the nonadiabatic couplings are
included or not (see the inset of the middle right-hand panel of
Fig.~\ref{fig:retinal_control})}. Because increasing the population of the
excited state has almost the same effect as increasing the molecular energy,
very similar dynamics and control pulses are obtained. Yet, the energy and
population controls do not always yield similar results. In the retinal model,
when performing energy control, no vibrational energy is pumped into the
system because the diagonal terms of the electric-dipole moment operator are
all zero by construction (hence $\langle\lbrack\hat{\boldsymbol{\mu}}%
,\hat{\mathbf{T}}]\rangle_{\psi_{t}}=0$). Consequently, only electronic
potential energy is added to the system, and the corresponding control pulse
is similar to the one obtained from the population control.

\begin{figure*}
[htb]%
\centering\includegraphics[width=0.95\linewidth]{
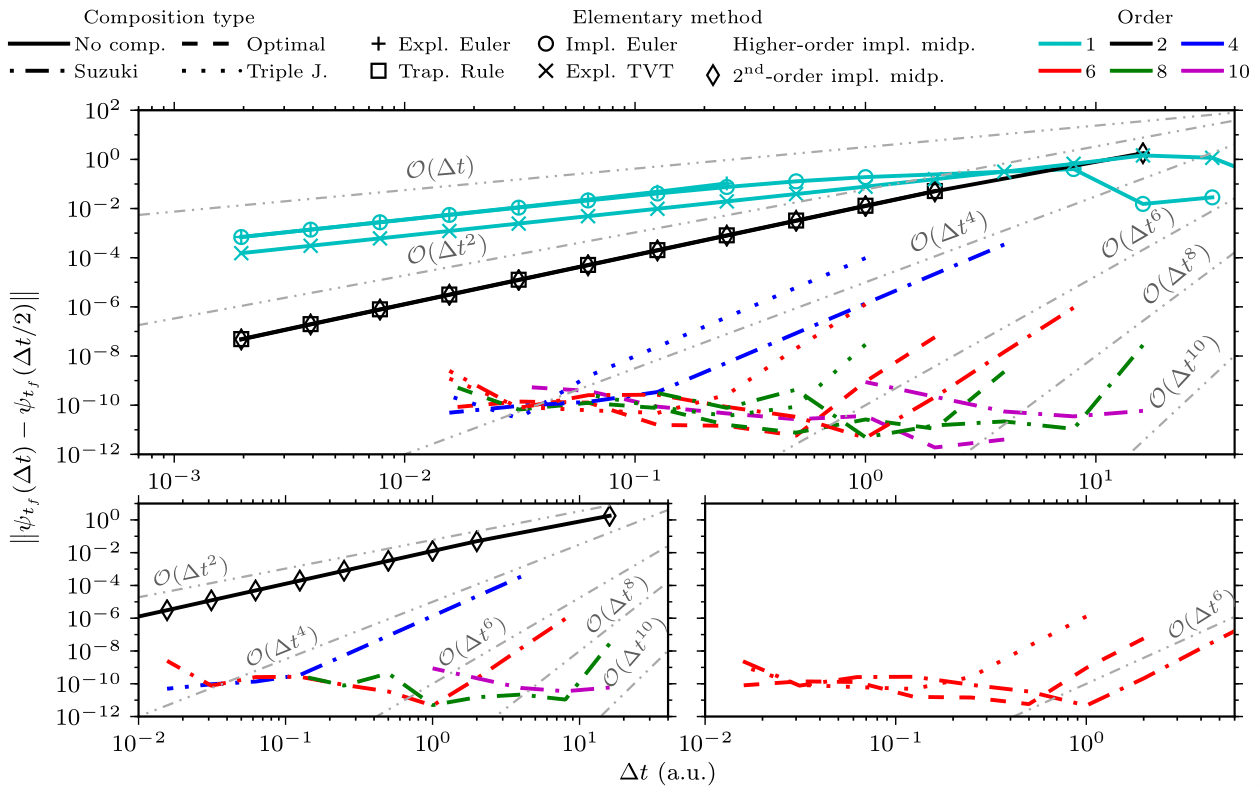}\caption{Convergence
of the molecular wavefunction at the final time $t_{f}$ achieved by the local
population control in the presence of nonadiabatic couplings. Top: All studied
methods, i.e., explicit and implicit Euler methods, {approximate
explicit TVT split-operator algorithm, {trapezoidal rule}}, implicit
midpoint method and its symmetric compositions. Bottom-left: Methods obtained
with the Suzuki composition. Bottom-right: Sixth-order methods obtained with
different composition schemes.}\label{fig:error_vs_dt_retinal_nacs_el_states}
\end{figure*}

\begin{figure}
[htb]%
\centering\includegraphics[width=0.95\linewidth]{
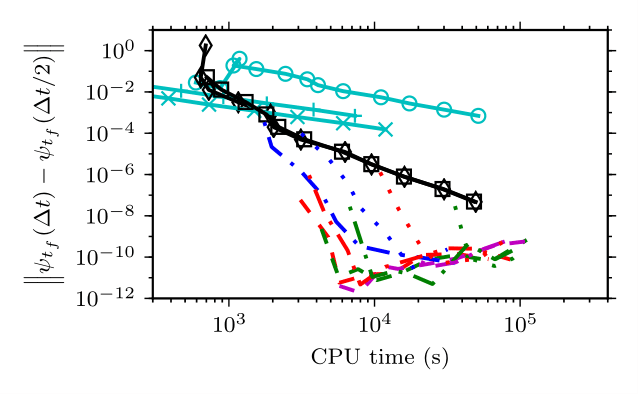}
\caption{Efficiency of the integrators used for the local population control of
retinal in the presence of nonadiabatic couplings. Efficiency is measured by plotting the convergence error
as a function of the computational (CPU) cost. Line labels are the same as in
Fig.~\ref{fig:error_vs_dt_retinal_nacs_el_states}.}
\label{fig:error_vs_CPU_time_retinal_nacs_el_states}
\end{figure}

To verify the orders of convergence predicted in
Sec.~\ref{subsec:lct_geom_methods}, we performed convergence analysis of
control simulations using various integrators. Simulations with each
integrator were repeated with different time steps and the resulting
wavefunctions at the final time $t_{f}$ were compared. As a measure of the
convergence error, we used the $L_{2}$-norm $\Vert\psi_{t_{f}}(\Delta
t)-\psi_{t_{f}}(\Delta t/2)\Vert$, where $\psi_{t}(\Delta t)$ denotes the
state at time $t$ obtained after propagation with time step $\Delta t$.
Figure~\ref{fig:error_vs_dt_retinal_nacs_el_states} displays the convergence
behavior of both Euler methods, {approximate explicit} TVT
split-operator algorithm, { trapezoidal rule}, and the proposed
implicit midpoint method as well as its symmetric compositions, when
controlling the excited state population. Notice that all integrators have
their predicted orders of convergence. The {approximate explicit}
TVT split-operator algorithm is, for the reasons mentioned in
Sec.~\ref{subsec:naive_so}, only first-order and not second-order as one might
na\"{\i}vely expect. For the convergence of other simulations, we refer the
reader to Figs.~S4-S6 of the supplementary material. Together, these results
confirm that both population and energy control follow the predicted order of
convergence regardless of the presence of nonadiabatic couplings.

\begin{figure}
[htb]%
\centering\includegraphics[width=0.95\linewidth]{
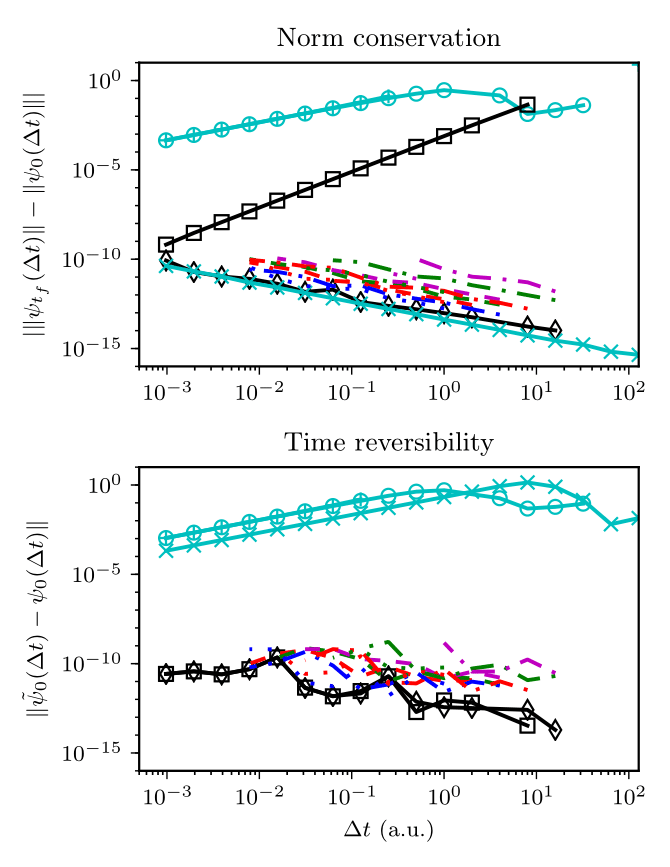}\caption{{Norm conservation 
(left) and time reversibility (right) of various integrators at the final time 
$t_{f}$
as a function of the time step $\Delta t$ used for the local control of population in the presence of
nonadiabatic couplings.
Time reversibility is measured by the distance between the initial state
$\psi_{0}$ and a ``forward-backward'' propagated state $\tilde{\psi}_{0}:=\hat{U}(0,t)\hat{U}(t,0)\psi_{0}$ [Eq.~(\ref{eq:forward_backward})] and line labels are
the same as in Fig.~\ref{fig:error_vs_dt_retinal_nacs_el_states}.}}\label{fig:geometric_properties_retinal_nacs_el_states}%

\end{figure}

\begin{figure}
[htb]%
\centering\includegraphics[width=0.95\linewidth]{
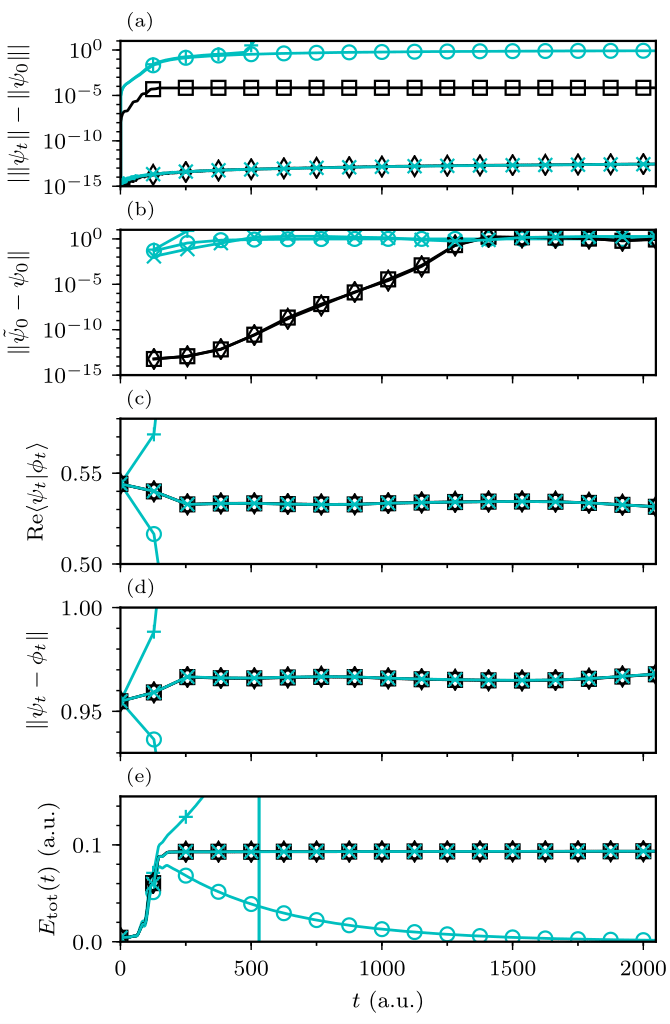}\caption{{Geometric 
properties of various integrators used for the local population control
in the presence of nonadiabatic couplings.
Panel~(a) shows that only the implicit midpoint and approximate explicit split-operator methods conserve
the norm, while panel~(b) demonstrates that only the implicit midpoint {method and the trapezoidal rule} are
time-reversible. (Reversibility is measured as in Fig.~\ref{fig:geometric_properties_retinal_nacs_el_states}).
Bottom three panels show that no method conserves (c) the inner product, (d) distance
between two states (which would imply stability), or (e) total energy $E_{\text{tot}}(t):=E_{0}(t) +
\langle\hat{\mathbf{V}}_{\text{LCT}}(\psi_{t})\rangle_{\psi_{t}}$
because even the exact evolution operator does not preserve these properties.
State $\phi_{0}$ is $\psi_{0}$ displaced along the reaction coordinate, i.e., a Gaussian wavepacket
with parameters $q_{0}=(0.1,0)$, $p_{0}=(0,0)$, and $\sigma_{0}= (0.128,0)$ a.u.
The time step $\Delta t= 2^{-2}$~a.u. was used for all calculations and line labels are
the same as in Fig.~\ref{fig:error_vs_dt_retinal_nacs_el_states}.
Note that only a few points of the Euler methods are visible in some of the plots because the results of
the Euler methods leave the range of these plots very rapidly. }}\label{fig:geometric_properties}%

\end{figure}

Because the higher-order methods require more work to perform each step, a
higher order of convergence may not guarantee higher efficiency. Therefore, we
evaluated the efficiency of each method directly by measuring the
computational cost needed to reach a prescribed convergence error.
Figure~\ref{fig:error_vs_CPU_time_retinal_nacs_el_states} shows the
convergence error as a function of the central processing unit (CPU) time and
confirms that, except for very crude calculations, higher-order integrators
are more efficient than any of the first- and second-order methods. For
example, to reach errors below a rather high threshold of {$3\times
10^{-4}$}, the fourth-order integrator obtained with the Suzuki composition
scheme is \emph{already} more efficient than any of the first- or second-order
algorithms. The efficiency gain increases further when highly accurate results
are desired. Indeed, for an error of $10^{-9}$, the eighth-order optimal
method is {$48$} times faster than the basic, second-order implicit
midpoint method and {approximately $400000$ times faster than the approximate explicit TVT split-operator algorithm (for which, due
to its inefficiency, the speedup had to be estimated by linear
extrapolation).} {High accuracy is hard to achieve with the
explicit methods because both the explicit Euler and approximate explicit TVT
split-operator algorithms have only first-order convergence.} Notice also that
the cost of implicit methods is not a monotonous function of the error because
the Newton-Raphson method needs more iterations to converge for larger than
for smaller time steps. Indeed, for time steps (or errors) larger than a
critical value, the CPU time might in fact increase with further increasing
time step (or error). The efficiency plots of other control simulations (see
Figs. S7-S9 of the supplementary material) confirm that the increase in
efficiency persists regardless of the control target (energy or
population)\ and presence or absence of nonadiabatic couplings.

{Figure~\ref{fig:geometric_properties_retinal_nacs_el_states}
analyzes how the time reversibility and norm conservation depend on the time
step. The figure confirms that all proposed integrators are exactly
time-reversible and norm-conserving regardless of the time step (the slow
increase of the error with decreasing time step is due to the accumulation of
numerical roundoff errors because a smaller time steps requires a larger
number of steps to reach the same final time $t_{f}$). In contrast, the figure
shows that an unrealistically small time step would be required for the
{trapezoidal rule} and both Euler methods to conserve norm exactly and for the
explicit split-operator algorithm to be exactly time-reversible.
Figures~S10-S12 of the supplementary material confirm that neither the chosen
control objective nor the nonadiabatic couplings influence the geometric
properties of the integrators.}

{We also checked how the conservation of geometric properties by
the integrators depend on time $t$ for a fixed time step. For these
simulations, we used a greater final time $t_{f}=2048$~a.u. and an
intentionally large time step $\Delta t=2^{-2}$~a.u. The grid was modified to
256 points between $\theta=\pm3\pi/2$~a.u. and 64 points between $q_{c}=\pm
9$~a.u., ensuring that the grid representation of the wavefunction at the
final time $t_{f}$ was converged (see Fig.~S2 of the supplementary material). Figure \ref{fig:geometric_properties} displays the
time evolution of the geometric properties for the elementary integrators
(i.e., {the trapezoidal rule, implicit midpoint,} approximate explicit
split-operator, and both Euler methods). The top panel confirms that the
implicit midpoint method and the approximate explicit TVT split-operator
algorithm conserve the norm exactly (i.e., to machine precision) even though a
large time step was used. In contrast, the {trapezoidal rule} and both Euler
methods do not conserve the norm. The second panel shows that only the
{ trapezoidal rule} and the implicit midpoint method are time-reversible.}
{However, due to the nonlinearity of the Schr\"{o}dinger equation
(\ref{eq:nl_TDSE}) and the accumulation of roundoff errors, the time
reversibility of these integrators slowly deteriorates as time increases. (For
a more detailed analysis of this gradual loss of time reversibility, we refer
the reader to Sec.~V of the supplementary material.) The bottom three
panels of Fig.~\ref{fig:geometric_properties} confirm that even
the implicit midpoint method does not conserve the inner product, distance
between two states, and total energy; this is not surprising
because, due to nonlinearity, the exact solution does not conserve these
properties either. Figures S13-S15 of the supplementary material also
confirm that neither the chosen control objective nor the nonadiabatic
couplings influence the time evolution of the geometric properties of these integrators.}

\section{Conclusion}

We presented high-order time-reversible integrators for the nonlinear
time-dependent Schr\"{o}dinger equation and demonstrated their efficiency and
geometric properties on the problem of local control of quantum systems. The
basic time-reversible integrator is an adaptation of the implicit midpoint
method to the nonlinear Schr\"{o}dinger equation and is obtained by composing
the explicit and implicit Euler methods. It is norm-conserving, symmetric,
time-reversible, and of second order of accuracy in the time step. Because it
is symmetric, the implicit midpoint method can be composed using symmetric
composition methods to obtain integrators of an arbitrary even order of
accuracy. These higher-order integrators conserve all of the properties of the
original second-order method.

In contrast, the {explicit} TVT split-operator algorithm is
generally only an approximate adaptation of the standard second-order TVT
split-operator algorithm to the nonlinear Schr\"{o}dinger equation
{which results from LCT.} Because this integrator {is
not implicit}, it is only of first order accuracy in the time step and loses
time reversibility while still conserving the norm. {The
{ trapezoidal rule}, another popular algorithm for solving the
Schr\"{o}dinger equation, remains symmetric and time-reversible, but does not
conserve the norm of the wavefunction propagated with a nonlinear
Schr\"{o}dinger equation.}

{Although we applied the proposed algorithms only to the special case of LCT,
they should be useful for any nonlinear time-dependent Schr\"{o}dinger
equation if high accuracy, norm conservation, and time
reversibility of the solution are desired.}

The data that support the findings of this study are contained in the paper
and supplementary material.

\section*{{Supplementary material}}

See the supplementary material for a detailed description of the generalized minimal residual
method, convergence of the wavefunction with respect to the grid density, a plot of the potential energy surfaces, an additional analysis of the convergence, efficiency, and geometric properties of various integrators, and a detailed analysis of the gradual loss of time reversibility of the implicit midpoint method.

\begin{acknowledgments}
The authors thank Seonghoon Choi for useful discussions and acknowledge the
financial support from the Swiss National Science Foundation within the
National Center of Competence in Research \textquotedblleft Molecular
Ultrafast Science and Technology\textquotedblright\ (MUST) and from the
European Research Council (ERC) under the European Union's Horizon 2020
research and innovation programme (grant agreement No. 683069 -- MOLEQULE).
\end{acknowledgments}

\appendix

\section{Geometric properties of various integrators}

\label{sec:proof_geometric_prop}

Here, we verify which geometric properties of the exact evolution are
preserved by various integrators. The analysis generalizes the analysis from
Appendix A of Ref.~\onlinecite{Choi_Vanicek:2019} for the linear to the
nonlinear Schr\"{o}dinger equation. To simplify the proofs, wherever it is not
ambiguous, we shall use abbreviated notation $\hat{U}_{\text{appr}}%
(\psi):=\hat{U}_{\text{appr}}(t+\Delta t,t;\psi)$ for the evolution operator
for a single time step and $\epsilon:=\Delta t/\hbar$ for the time step
divided by Planck's constant.

\subsection{Norm conservation}

In general, the norm is conserved if and only if
\[
\hat{U}_{\text{appr}}(\psi)^{\dagger}\hat{U}_{\text{appr}}(\psi)=1,
\]
which follows from a derivation analogous to Eq.~(\ref{eq:exact_norm}) for the
exact operator $\hat{U}(\psi)$. As in the linear case, neither Euler method
conserves the norm because
\begin{align}
\hat{U}_{\text{expl}}(\psi_{t})^{\dagger}\hat{U}_{\text{expl}}(\psi_{t})  &
=[1+i\epsilon\hat{H}(\psi_{t})][1-i\epsilon\hat{H}(\psi_{t})]\nonumber\\
&  =1+\epsilon^{2}\hat{H}(\psi_{t})^{2}\neq1 \label{eq:norm_expl}%
\end{align}
and
\begin{align}
&  \hat{U}_{\text{impl}}(\psi_{t+\Delta t})^{\dagger}\hat{U}_{\text{impl}%
}(\psi_{t+\Delta t})\nonumber\\
&  \qquad=[1-i\epsilon\hat{H}(\psi_{t+\Delta t})]^{-1}[1+i\epsilon\hat{H}%
(\psi_{t+\Delta t})]^{-1}\nonumber\\
&  \qquad=[1+\epsilon^{2}\hat{H}(\psi_{t+\Delta t})^{2}]^{-1}\neq1.
\label{eq:norm_impl}%
\end{align}

The trapezoidal rule, norm-conserving in the linear case, loses this property
for nonlinear Hamiltonians since
\begin{align}
&  \hat{U}_{\text{trap}}(\psi)^{\dagger}\hat{U}_{\text{trap}}(\psi)\nonumber\\
&  \quad=[1+i\epsilon\hat{H}(\psi_{t})/2][1-i\epsilon\hat{H}(\psi_{t+\Delta
t})/2]^{-1}\nonumber\\
&  \qquad\times\lbrack1+i\epsilon\hat{H}(\psi_{t+\Delta t})/2]^{-1}%
[1-i\epsilon\hat{H}(\psi_{t})/2]\nonumber\\
&  \quad=[1+i\epsilon\hat{H}(\psi_{t})/2][1+\epsilon^{2}\hat{H}(\psi_{t+\Delta
t})^{2}/4]^{-1}\nonumber\\
&  \qquad\times\lbrack1-i\epsilon\hat{H}(\psi_{t})/2]\nonumber\\
&  \quad\neq1;
\end{align}
the last nontrivial expression does not reduce to $1$ {because }$\hat{H}%
(\psi_{t})\neq\hat{H}(\psi_{t+\Delta t})$. {In contrast, both the implicit
midpoint and approximate explicit TVT split-operator algorithms conserve the
norm even in the nonlinear setting because}
\begin{align}
&  \hat{U}_{\text{mid}}(\psi_{t+\Delta t/2})^{\dagger}\hat{U}_{\text{mid}%
}(\psi_{t+\Delta t/2})\nonumber\\
&  \quad=[1-i\epsilon\hat{H}(\psi_{t+\Delta t/2})/2]^{-1}[1+i\epsilon\hat
{H}(\psi_{t+\Delta t/2})/2]\nonumber\\
&  \qquad\times\lbrack1-i\epsilon\hat{H}(\psi_{t+\Delta t/2})/2][1+i\epsilon
\hat{H}(\psi_{t+\Delta t/2})/2]^{-1}\nonumber\\
&  \quad=1
\end{align}
(where in the last step we used the commutativity of the middle two factors in
the previous expression) and
\begin{align}
&  \hat{U}_{\text{expl TVT}}({\psi_{t,\hat{T}\Delta t/2}})^{\dagger}\hat
{U}_{\text{expl TVT}}({\psi_{t,\hat{T}\Delta t/2}})\nonumber\\
&  \qquad=e^{i\epsilon\hat{T}/2}e^{i\epsilon\hat{V}_{\text{tot}}({\psi
_{t,\hat{T}\Delta t/2}})}e^{i\epsilon\hat{T}/2}\nonumber\\
&  \quad\qquad\times e^{-i\epsilon\hat{T}/2}e^{-i\epsilon\hat{V}_{\text{tot}%
}({\psi_{t,\hat{T}\Delta t/2}})}e^{-i\epsilon\hat{T}/2}\nonumber\\
&  \qquad=1.
\end{align}

\subsection{Symmetry and time reversibility}

Neither Euler method is symmetric or time-reversible because they are adjoints
of each other:
\begin{align}
\hat{U}_{\text{expl}}(t+\Delta t,t;\psi_{t})^{\ast}  &  =\hat{U}_{\text{expl}%
}(t,t+\Delta t;\psi_{t+\Delta t})^{-1}\nonumber\\
&  =[1-i(-\Delta t)\hat{H}(\psi_{t+\Delta t})/\hbar]^{-1}\nonumber\\
&  =\hat{U}_{\text{impl}}(t+\Delta t,t;\psi_{t+\Delta t}).
\end{align}
The approximate explicit TVT split-operator algorithm is also
time-irreversible because forward propagation is not cancelled by backward
propagation:
\begin{align}
&  \hat{U}_{\text{expl TVT}}(t;t+\Delta t,\psi_{\hat{V}\hat{T}})\hat
{U}_{\text{expl TVT}}(t+\Delta t;t,{\psi_{t,\hat{T}\Delta t/2}})\nonumber\\
&  \qquad=e^{i\epsilon\hat{T}/2}e^{i\epsilon\hat{V}_{\text{tot}}(\psi_{\hat
{V}\hat{T}})}e^{i\epsilon\hat{T}/2}\nonumber\\
&  \quad\qquad\times e^{-i\epsilon\hat{T}/2}e^{-i\epsilon\hat{V}_{\text{tot}%
}({\psi_{t,\hat{T}\Delta t/2}})}e^{-i\epsilon\hat{T}/2}\nonumber\\
&  \qquad\neq1, \label{eq:naive_VTV_time_revers}%
\end{align}
where ${\psi_{\hat{V}\hat{T}}:=e^{i\epsilon\hat{T}/2}\psi_{t+\Delta
t}=e^{-i\epsilon\hat{V}_{\text{tot}}(\psi_{t,\hat{T}\Delta t/2})}}$%
{$\psi_{t,\hat{T}\Delta t/2}$ denotes the state obtained after forward
propagation of $\psi_{t,\hat{T}\Delta t/2}$ with the potential evolution
operator and $\hat{V}_{\text{tot}}(\psi_{\hat{V}\hat{T}})\neq\hat
{V}_{\text{tot}}(\psi_{t,\hat{T}\Delta t/2})$ was used to obtain the last
line. It is clear from Eq.~(\ref{eq:naive_VTV_time_revers}) that if the
nonlinear term is as in the Gross-Pitaevskii equation, i.e., if $V_{\text{tot}%
}(\psi,q)=V_{0}(q)+C|\psi(q)|^{2}$ with }$C${ a real constant, the approximate
explicit TVT split-operator algorithm becomes time-reversible because $\hat
{V}_{\text{tot}}(\psi_{\hat{V}\hat{T}})=\hat{V}_{\text{tot}}(\psi_{t,\hat
{T}\Delta t/2})$ in that particular case; the reason is that wavefunctions
$\psi_{\hat{V}\hat{T}}(q)$ and $\psi_{t,\hat{T}\Delta t/2}(q)$ in position
representation only differ by a position-dependent phase factor. However, the
equality $\hat{V}_{\text{tot}}(\psi_{\hat{V}\hat{T}})=\hat{V}_{\text{tot}%
}(\psi_{t,\hat{T}\Delta t/2})$ does not hold for other nonlinear Hamiltonians,
such as the one in LCT, which contains a more general nonlinearity. }

In contrast, both {the implicit midpoint method and trapezoidal rule are
always time-reversible because they are symmetric:}%
\begin{align}
&  \hat{U}_{\text{mid}}(t+\Delta t,t;\psi_{t+\Delta t/2})^{\ast}=\hat
{U}_{\text{mid}}(t,t+\Delta t;\psi_{t+\Delta t/2})^{-1}\nonumber\\
&  \quad=\left(  [1+i\epsilon\hat{H}(\psi_{t+\Delta t/2})/2][1-i\epsilon
\hat{H}(\psi_{t+\Delta t/2})/2]^{-1}\right)  ^{-1}\nonumber\\
&  \quad=[1-i\epsilon\hat{H}(\psi_{t+\Delta t/2})/2][1+i\epsilon\hat{H}%
(\psi_{t+\Delta t/2})/2]^{-1}\nonumber\\
&  \quad=\hat{U}_{\text{mid}}(t+\Delta t,t;\psi_{t+\Delta t/2}),
\end{align}
and
\begin{align}
&  \hat{U}_{\text{trap}}(t+\Delta t,t;\psi)^{\ast}=\hat{U}_{\text{trap}%
}(t,t+\Delta t;\psi)^{-1}\nonumber\\
&  \quad=\left(  [1-i\epsilon\hat{H}(\psi_{t})/2]^{-1}[1+i\epsilon\hat{H}%
(\psi_{t+\Delta t})/2]\right)  ^{-1}\nonumber\\
&  \quad=[1+i\epsilon\hat{H}(\psi_{t+\Delta t})]^{-1}[1-i\epsilon\hat{H}%
(\psi_{t})/2]\nonumber\\
&  \quad=\hat{U}_{\text{trap}}(t+\Delta t,t;\psi).
\end{align}

\bibliographystyle{aipnum4-2}
\bibliography{LCT_arbitrary_order}

\end{document}


\title{Supplementary material: Time-reversible and norm-conserving high-order integrators for the nonlinear
time-dependent Schr\"{o}dinger equation:\ Application to local control theory}
\author{Julien Roulet}
\email{julien.roulet@epfl.ch}
\author{Ji{\v r}{\'i} Van{\'i}{\v c}ek}
\email{jiri.vanicek@epfl.ch}
\affiliation{Laboratory of theoretical physical chemistry, Institut des sciences et
ing{\'e}nieries Chimiques, Ecole Polytechnique F\'ed\'erale de Lausanne (EPFL), Lausanne, Switzerland}
\date{\today}
\maketitle

\setcounter{table}{0} \renewcommand{\thetable}{S\arabic{table}}
\setcounter{figure}{0} \renewcommand{\thefigure}{S\arabic{figure}}

\section{{The generalized minimal residual method}}

\label{sec:GMRES}

The generalized minimal residual method is an iterative method for solving
linear systems
\begin{equation}
Ax=b,
\end{equation}
such as the one encountered in Eq.~(33), with $A\equiv J(\psi^{(k)})$,
$x\equiv\delta\psi^{(k)}$, and $b\equiv-f(\psi^{(k)})$. Here, we used the
restart version of the algorithm. Following Refs.~\onlinecite{book_Saad:2003}
and \onlinecite{Saad_Schultz:1986}, the algorithm can be summarized in four steps:

\begin{enumerate}
\item Initialization

\item Arnoldi's method for constructing an orthonormal basis of the Krylov subspace

\item Forming the solution\label{item:form_solution}

\item Restart
\end{enumerate}

Each step is described in more detail below.

\subsection{Initialization}

The initialization is performed as follows: Choose an initial guess $x_{0}$,
an error threshold $\epsilon$, a maximal number of iterations $m$, and compute
the initial residue $r_{0}=b-Ax_{0}$, $\beta:=\Vert r_{0}\Vert$, and
$v_{1}:=r_{0}/\beta$.

\subsection{Arnoldi's method}

Arnoldi's method obtains an orthonormal basis of the Krylov
subspace\cite{Arnoldi:1951,Saad:1980}%
\[
{\mathcal{K}_{m}:=\text{span}\{v_{1},Av_{1},\dots,A^{m-1}v_{1}\}.}%
\]
{The orthonormal vectors, which span the Krylov subspace, are obtained using
the modified Gram-Schmidt process. The algorithm is: }

\begin{enumerate}
\item For $j=1,2,\dots,m$ Do: {\setlength\itemindent{2em} }

\item Compute $w_{j}:=Av_{j}$ {\setlength\itemindent{2em}}

\item For $i=1,\dots j$ Do: {\setlength\itemindent{4em}}

\item $h_{ij}:=\langle v_{i}|w_{j}\rangle$\label{item:h_ij}
{\setlength\itemindent{4em}}

\item $w_{j}:=w_{j}-h_{ij}v_{i}$ {\setlength\itemindent{2em} }

\item End Do {\setlength\itemindent{2em}}

\item $h_{j+1,j} = \|w_{j}\|$. If $h_{j+1,j} =0$, set $m=j$ and stop
\label{item:h_jj+1} {\setlength\itemindent{2em}}

\item $v_{j+1}= w_{j}/h_{j+1,j}$

\item End Do
\end{enumerate}

\subsection{Forming the solution}

Here, the goal is to find the solution which minimizes the norm of
the residue after $m$ iterations of Arnoldi's method. Now that the orthogonal
basis is constructed, we know that the solution lies in the affine subspace
$x_{0}+\mathcal{K}_{m}$. Thus, any vector $x$ in that space can be written
as
\begin{equation}
x=x_{0}+V_{m}y,
\end{equation}
where $V_{m}:=\{v_{1},\dots,v_{m}\}$ is the matrix constructed
from the orthonormal vectors $v_{j}$ and $y$ is an arbitrary vector of size
$m$. The problem translates to the minimization of the function
\begin{align}
M(y)  &  =\Vert b-A(x_{0}+V_{m}y)\Vert\nonumber\\
&  =\Vert r_{0}-AV_{m}y\Vert\nonumber\\
&  =\Vert\beta v_{1}-V_{m+1}\bar{H}_{m}y\Vert\nonumber\\
&  =\Vert V_{m+1}(\beta e_{1}-\bar{H}_{m}y)\Vert\nonumber\\
&  =\Vert(\beta e_{1}-\bar{H}_{m}y)\Vert, \label{eq:min_problem}%
\end{align}
where $\bar{H}_{m}=\{h_{ij}\}_{1\leq i\leq m+1,1\leq j\leq m}$ is
the the $(m+1)\times m$ Hessenberg matrix constructed from the elements
$h_{ij}$ obtained in Arnoldi's method and where $e_{1}=(1,0,\dots,0)^{T}$ is
the first vector in the standard basis of $\mathbb{R}^{m+1}$. The third line
of Eq.~(\ref{eq:min_problem}) was obtained using $AV_{m}=V_{m+1}\bar{H}_{m}$,
while the last line follows because the columns of $V_{m+1}$ are orthonormal.
This step of the generalized minimal residual algorithm can be summarized as:

\begin{enumerate}
\item Construct the Hessenberg matrix $\bar{H}_{m}$\label{item:hessenberg}

\item Compute $y_{m}$, the vector of dimension $m$ minimizing $M(y)$

\item Compute the solution $x_{m}= x_{0}+V_{m}y_{m}$
\end{enumerate}

\subsection{Restart}

This last step of the generalized minimal residual algorithm
determines whether the solution $x_{m}$, obtained at the previous step, is
sufficiently accurate. If the condition $\Vert r_{m}\Vert:=\Vert b-Ax_{m}%
\Vert\leq\epsilon$ holds, $x_{m}$ is taken as the solution. If $\Vert
r_{m}\Vert>\epsilon$, we set $x_{0}=x_{m}$ and restart the algorithm.

\subsection{Practical implementation}

In practice, it is difficult to compute the vector $y_{m}$ minimizing $M(y)$.
For this, it is preferable to use the decomposition
\begin{equation}
\bar{R}_{m}=Q_{m}\bar{H}_{m}, \label{eq:QR}%
\end{equation}
where $Q_{m}$ is an $(m+1)\times(m+1)$ orthogonal matrix and $\bar{R}_{m}$ is
an upper triangular matrix of dimension $(m+1)\times m$ with a bottom row
consisting only of zeroes, i.e., it can be decomposed as
\begin{equation}
\bar{R}_{m}=%
\begin{pmatrix}
R_{m}\\
0
\end{pmatrix}
,
\end{equation}
where $R_{m}$ is an $m\times m$ upper triangular matrix. Once this
factorization is performed, the minimization problem simplifies to
\begin{equation}
y_{m}=\operatorname{argmin}_{y}\Vert\beta e_{1}-\bar{H}_{m}y\Vert
=\operatorname{argmin}_{y}\Vert\bar{g}_{m}-\bar{R}_{m}y\Vert,
\end{equation}
where $\bar{g}_{m}:=Q_{m}\beta\epsilon_{1}=(\gamma_{1},\dots,\gamma_{m+1}%
)^{T}$ and can be decomposed as $\bar{g}_{m}=(g_{m},\gamma_{m+1})^{T}$. The
solution of the minimization problem is now trivially obtained by solving the
linear triangular system $R_{m}y_{m}=g_{m}$, which yields
\begin{equation}
y_{m}=R_{m}^{-1}g_{m}.
\end{equation}
Note that it can be proven that the residue after $m$ iterations of the
algorithm is given by the last element of $\bar{g}_{m}$, i.e., $r_{m}%
=|\gamma_{m+1}|$.\cite{book_Saad:2003}

Instead of constructing $\bar{H}_{m}$ and computing its decomposition
(\ref{eq:QR}) after $m$ iterations of Arnoldi's method, both can be performed
in an iterative manner, at each iteration $j$ of Arnoldi's method. This makes
it possible to obtain $\bar{g}_{j}$ and therefore to check if $r_{j}%
\leq\epsilon$ at each iteration step $j$, without even solving the
minimization problem.

To do so, $\bar{H}_{j+1}$ is obtained by concatenation of $\bar{H}_{j}$ and
the elements obtained in steps \ref{item:h_ij} and \ref{item:h_jj+1} of
Arnoldi's method, i.e.,
\begin{equation}
\bar{H}_{j+1}=%
\begin{pmatrix}
& h_{1,j+1}\\
\bar{H}_{j} & \vdots\\
& h_{j+1,j+1}\\
0 & h_{j+2,j+1}%
\end{pmatrix}
.
\end{equation}
Then, multiplying $\bar{H}_{j+1}$ on the left by the orthogonal matrix
obtained at the previous iteration yields an almost upper triangular matrix%

\begin{equation}
\tilde{R}_{j+1}=%
\begin{pmatrix}
Q_{j} & 0\\
0 & 1
\end{pmatrix}
\bar{H}_{j+1}=%
\begin{pmatrix}
& r_{1,j+1}\\
R_{j} & \vdots\\
& r_{j,j+1}\\
0 & \eta\\
0 & \nu
\end{pmatrix}
\end{equation}
of dimension $(j+2)\times(j+1)$. Note that $\eta$ and $\nu$ denote
the arbitrary elements $j+1,j+1$ and $j+2,j+1$ of the matrix $\tilde{R}_{j+1}%
$. From there, $\bar{R}_{j+1}$, which is upper triangular, is obtained by
rotating $\tilde{R}_{j+1}$:
\begin{equation}
\bar{R}_{j+1}=Q_{j+1}\bar{H}_{j+1}=\Omega_{j}\tilde{R}_{j+1}=%
\begin{pmatrix}
& r_{1,j+1}\\
R_{j} & \vdots\\
0 & r_{j+1,j+1}\\
0 & 0
\end{pmatrix}
, \label{eq:R_j_plus_one}%
\end{equation}
where
\begin{equation}
\Omega_{j}=%
\begin{pmatrix}
1 &  &  &  & \\
& \ddots &  &  & \\
&  & 1 &  & \\
&  &  & c_{j}^{\ast} & s_{j}^{\ast}\\
&  &  & -s_{j} & c_{j}%
\end{pmatrix}
\label{eq:givens_rot}%
\end{equation}
is a rotation matrix of dimension $(j+2)\times(j+2)$. The complex
coefficients
\begin{equation}
c_{j}:=\frac{\eta}{\sqrt{|\eta|^{2}+\left\vert \nu\right\vert ^{2}}}%
\quad\text{ and }\quad s_{j}:=\frac{\nu}{\sqrt{|\eta|^{2}+\left\vert
\nu\right\vert ^{2}}}%
\end{equation}
of $\Omega_{j}$ are selected so that multiplying $\Omega_{j}$ with $\tilde
{R}_{j+1}$ yields the upper triangular matrix $\bar{R}_{j+1}$. From
Eq.~(\ref{eq:R_j_plus_one}), $Q_{j+1}$ is obtained by applying the rotation
matrix $\Omega_{j}$ to $Q_{j}$ as
\begin{equation}
Q_{j+1}=\Omega_{j}%
\begin{pmatrix}
Q_{j} & 0\\
0 & 1
\end{pmatrix}
.
\end{equation}

\section{Exponential convergence of the wavefunction with respect
to the grid density}

\begin{figure}
[htb!]\centering
\includegraphics[width=0.7\linewidth]{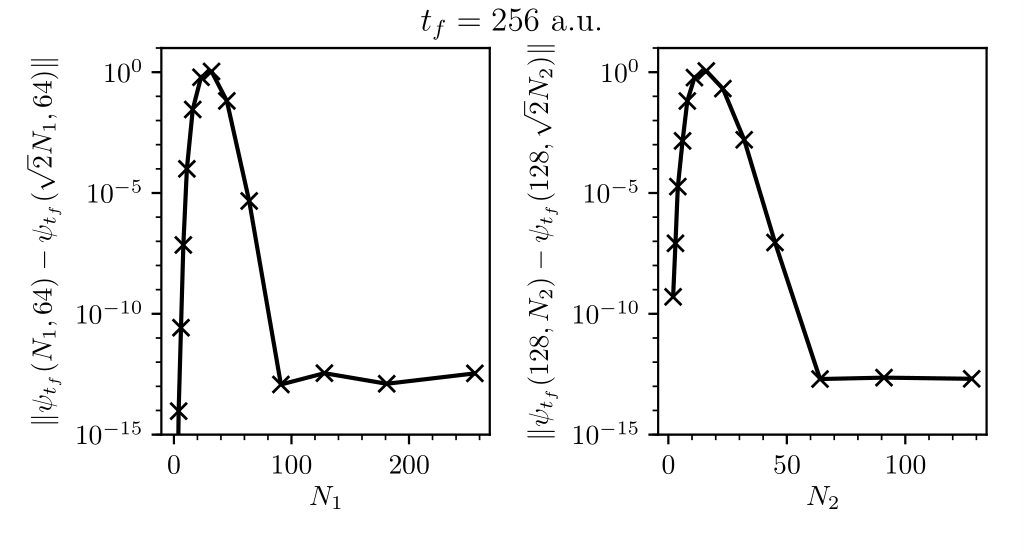} \caption{
Error of the wavefunction at the final time $t_{f}=256$ a.u. as a function of the number of grid points.
The figure indicates that the numerical grids consisting of $N_{1}=128$ points between $\theta=\pm \pi/2$ a.u.
and $N_{2}=64$ points between $q_{c} = \pm 9.0$ a.u. are converged. All calculations were performed using
the approximate explicit TVT split-operator algorithm with time step $\Delta t=2^{-1}$ a.u. }
\label{fig:grid_conv_256}
\end{figure}

\begin{figure}
[htb!]\centering\includegraphics[width=0.7\linewidth]{
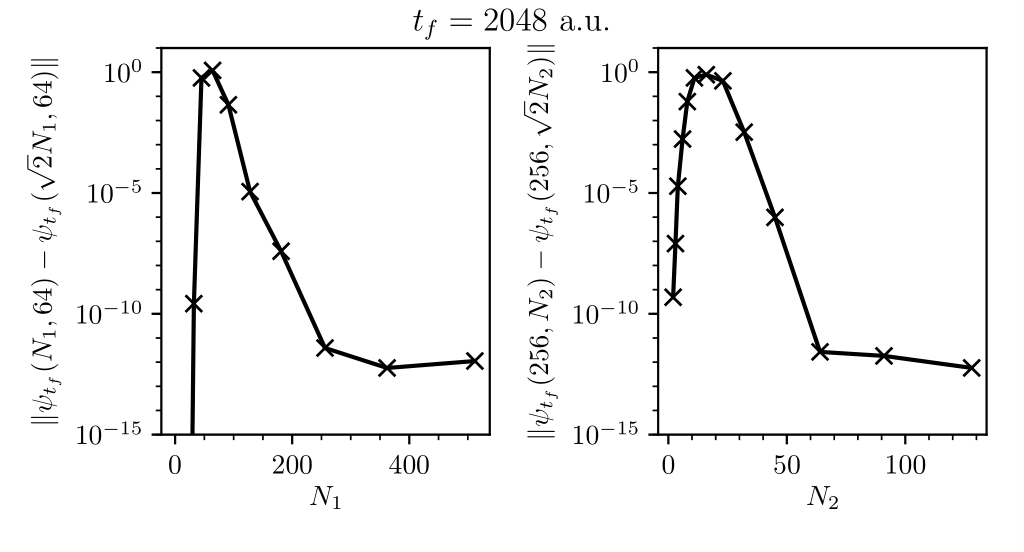}
\caption{Error of the wavefunction at the final time $t_{f}=2048$ a.u. as a function of the number of grid
points. The figure indicates that the numerical grids consisting of $N_{1}=256$ points between
$\theta=\pm 3\pi/2$ a.u. and $N_{2}=64$ points between $q_{c} = \pm 9.0$ a.u. are converged.
All calculations were performed using the approximate explicit TVT split-operator algorithm with time step
$\Delta t=2^{-1}$ a.u.} \label{fig:grid_conv_2048}
\end{figure}

Figures \ref{fig:grid_conv_256} and \ref{fig:grid_conv_2048} show the
convergence of the wavefunction with respect to the grid density at the final
times $t_{f}=256$ a.u. and $t_{f}=2048$ a.u., respectively. For both figures,
the goal of the calculation was to increase the population of the excited
state in the presence of nonadiabatic couplings and the approximate explicit
TVT split-operator algorithm was used with a time step $\Delta t=2^{-1}$ a.u.
The grid convergence was obtained by fixing the number of grid points of one
of the two dimensions and repeating the calculation with a varying number of
grid points for the other dimension. For each increase in the number of points
by a factor of $k\approx\sqrt{2}$ in the varying dimension, the
position and momentum ranges as well as densities in that dimension were
increased by a factor $\sqrt{k}\approx2^{1/4}$. Then, the error was measured
by the distance $\Vert\psi_{t_{f}}(N_{1},Y)-\psi_{t_{f}}(\sqrt{2}%
N_{1},Y)\Vert$ and $\Vert\psi_{t_{f}}(X,N_{2})-\psi_{t_{f}}(X,\sqrt{2}%
N_{2})\Vert$ for the grid convergence of the first and second dimension,
respectively, with $\psi_{t_{f}}(N_{1},N_{2})$ denoting the wavefunction at
the final time $t_{f}$, obtained using grids consisting of $N_{1}$ and $N_{2}$
points in the first and second dimension, respectively. Note also that $X$ and
$Y$ are the fixed numbers of grid points of the
first and second dimensions, respectively. The comparison of wavefunctions
that are represented on different grids was performed by trigonometric
interpolation of the wavefunction represented on the sparser grid.
In both figures, exponential convergence with respect to the grid
density of each dimension is observed.

\section{Potential energy surfaces}

\begin{figure}
[htb!]\centering
\includegraphics[width=0.5\linewidth]{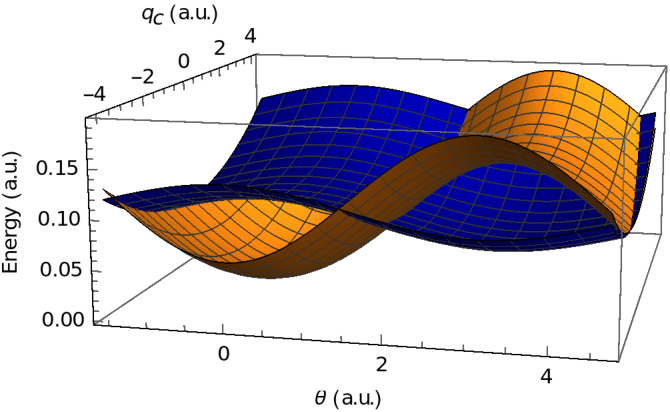}
\caption{Diabatic potential energy surfaces of the model.  Orange: Ground electronic state [Eq.~(42)]. Blue: Excited electronic state [Eq.~(43)].}
\label{fig:PES}
\end{figure}

Figure~\ref{fig:PES} displays the two diabatic potential energy surfaces.

\section{Convergence, efficiency, and geometric properties of the integrators
for the other simulations}

\begin{figure}
[htb!]\centering
\includegraphics[width=0.9\linewidth]{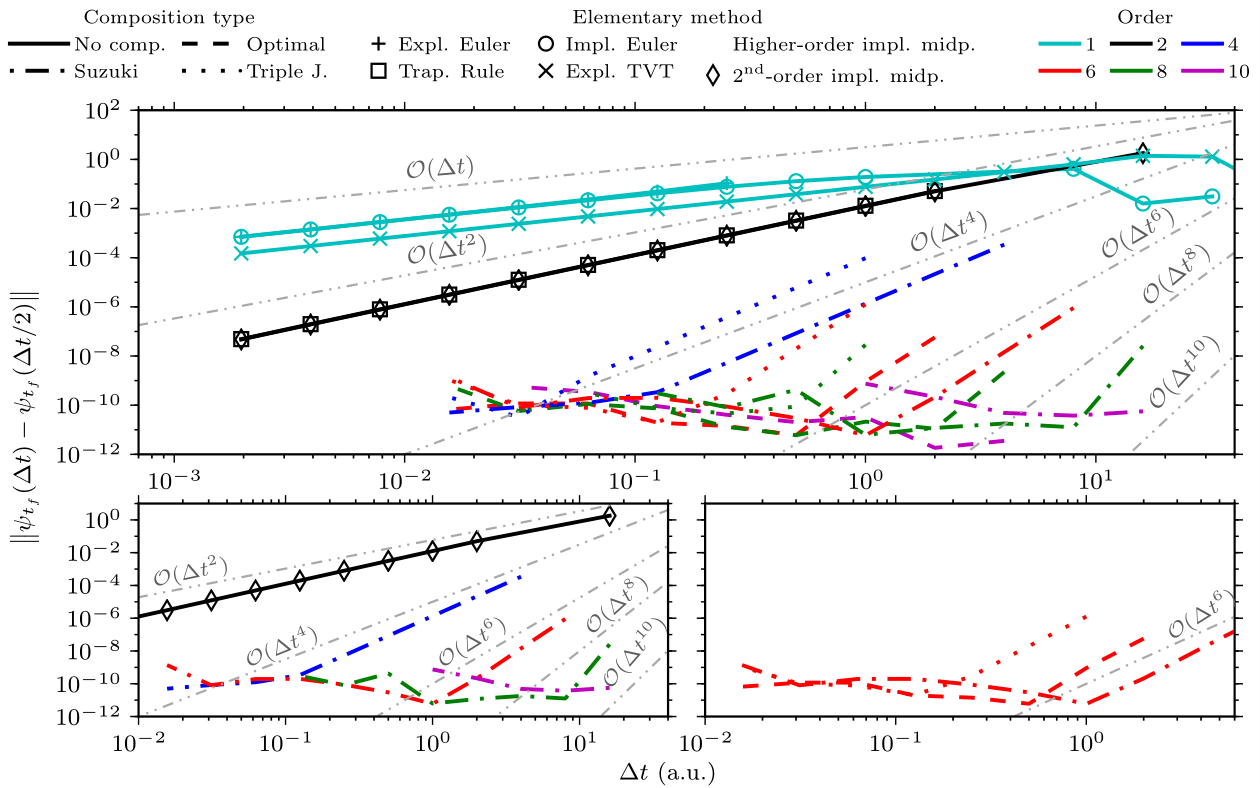}
\caption{Convergence
of the molecular wavefunction at the final time $t_{f}$ achieved by the local population
control in the absence of nonadiabatic couplings. Top: All
studied methods, i.e., explicit and implicit Euler methods, approximate explicit TVT split-operator
algorithm, {trapezoidal rule}, implicit midpoint method and its symmetric compositions.
Bottom-left: Methods obtained with the Suzuki composition.
Bottom-right: Sixth-order methods obtained with different composition
schemes.} \label{fig:error_vs_dt_nonacs_el_states}
\end{figure}

\begin{figure}
[htb!]\centering
\includegraphics[width=0.9\linewidth]{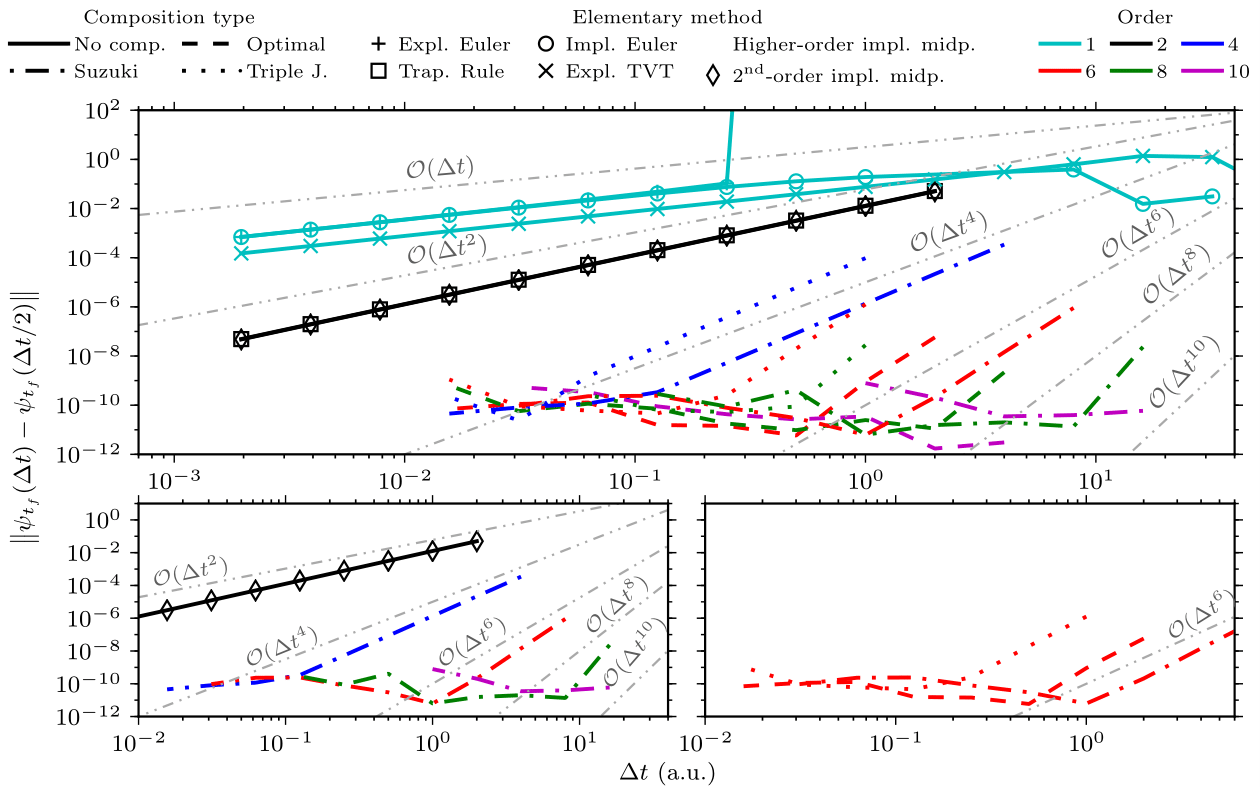}
\caption{Same as Fig.~\ref{fig:error_vs_dt_nonacs_el_states}, but for energy control without nonadiabatic couplings.}\label{fig:error_vs_dt_nonacs_energy}%

\end{figure}

\begin{figure}
[htb!]\centering
\includegraphics[width=0.5\linewidth]{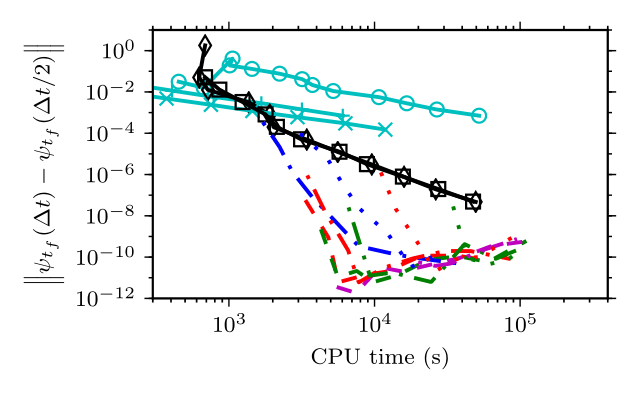}
\caption{Efficiency of the integrators used for the local population control of
retinal in the absence of nonadiabatic couplings. Efficiency is measured by plotting the convergence error
as a function of the computational (CPU) cost.}
\label{fig:error_vs_CPU_time_nonacs_el_states}
\end{figure}

\begin{figure}
[htb!]\centering
\includegraphics[width=0.5\linewidth]{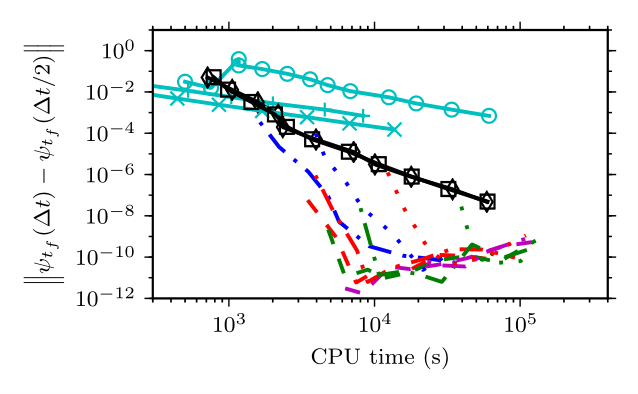}
\caption{Same as Fig.~\ref{fig:error_vs_CPU_time_nonacs_el_states}, but for energy control without nonadiabatic couplings.}\label{fig:error_vs_CPU_time_nonacs_energy}%

\end{figure}

\begin{figure}
[htb!]\centering
\includegraphics[width=0.5\linewidth]{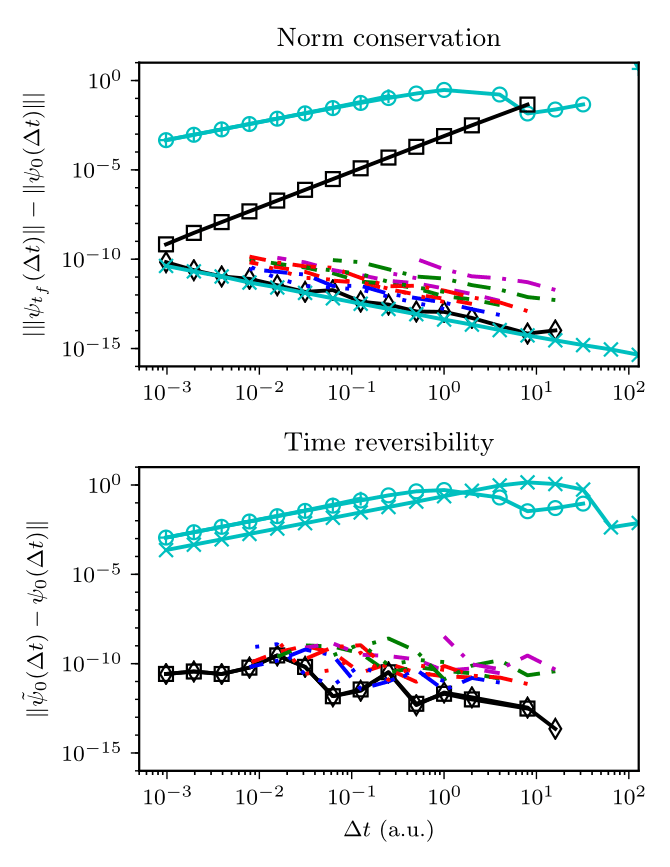}
\caption{Norm conservation (left) and time reversibility (right) of various integrators at the final time $t_{f}$
as a function of the time step $\Delta t$ used for the local population control in the absence of
nonadiabatic couplings. Time reversibility is measured by the distance between the initial state
$\psi_{0}^{\Delta t}$ and a ``forward-backward'' propagated state $\tilde{\psi}_{0}^{\Delta t}:=\hat{U}(0,t)\hat{U}(t,0)\psi_{0}^{\Delta t}$ [Eq.~(19)].}
\label{fig:geometric_properties_nonacs_el_states}
\end{figure}

We present here the results obtained when controlling the population (in the
absence of nonadiabatic couplings) as well as the energy (in the presence and
the absence of nonadiabatic couplings).
Figures~\ref{fig:error_vs_dt_nonacs_el_states}%
-\ref{fig:error_vs_dt_nonacs_energy} show the {convergence with
respect to the time step }$\Delta t$ of the wavefunction at the final time
$t_{f}$ for the different simulations and {Figs.}%
~\ref{fig:error_vs_CPU_time_nonacs_el_states}%
-\ref{fig:error_vs_CPU_time_nonacs_energy} {show the efficiency of
various algorithms}. {The geometric} properties are displayed in
Figs.~\ref{fig:geometric_properties_nonacs_el_states}%
-\ref{fig:nonconserved_geometric_properties_nonacs_energy}.

\begin{figure}
[htb!]\centering
\includegraphics[width=0.9\linewidth]{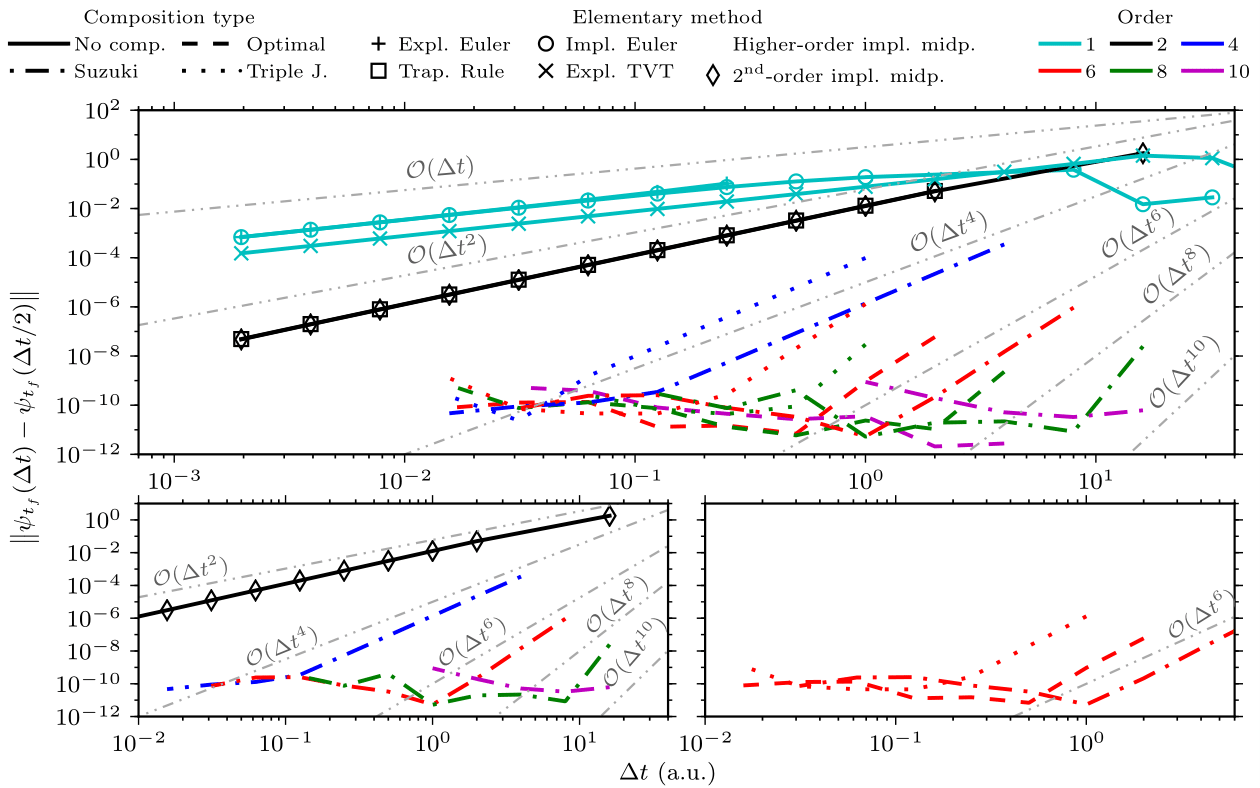}
\caption{Same as Fig.~\ref{fig:error_vs_dt_nonacs_el_states}, but for energy control with nonadiabatic couplings. }\label{fig:error_vs_dt_nacs_energy}%

\end{figure}

\begin{figure}
[htb!]\centering
\includegraphics[width=0.5\linewidth]{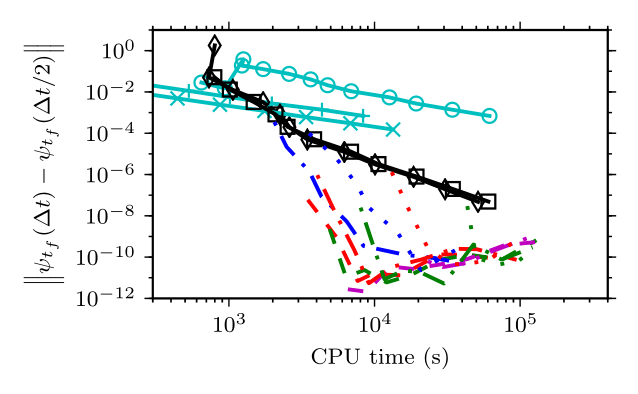}
\caption{Same as Fig.~\ref{fig:error_vs_CPU_time_nonacs_el_states}, but for energy control with nonadiabatic couplings.}\label{fig:error_vs_CPU_time_nacs_energy}%

\end{figure}

\begin{figure}
[htb!]\centering
\includegraphics[width=0.5\linewidth]{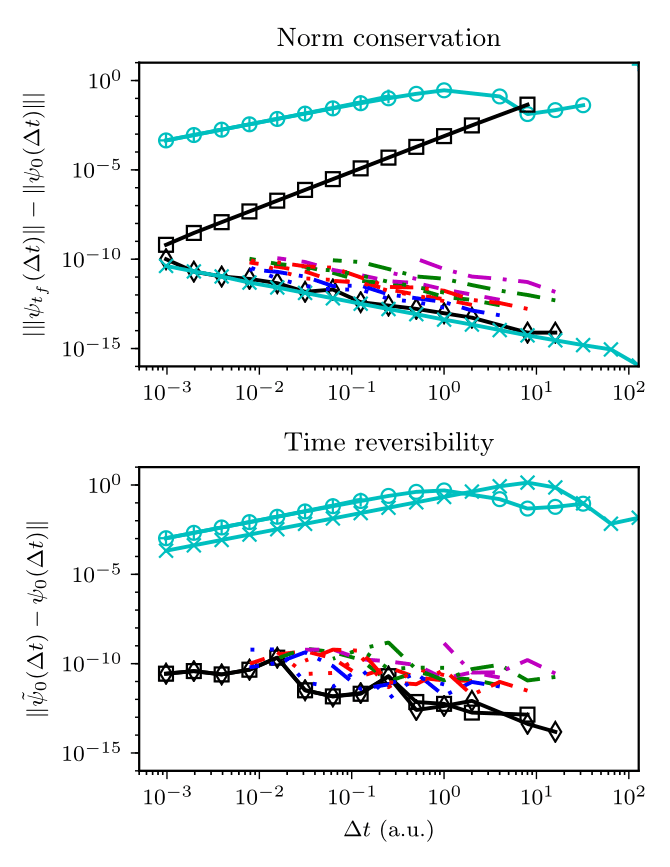}
\caption{Same as Fig~\ref{fig:geometric_properties_nonacs_el_states}, but for energy control in the presence of nonadiabatic couplings.}\label{fig:geometric_properties_nacs_energy}%

\end{figure}

\begin{figure}
[htb!]\centering
\includegraphics[width=0.5\linewidth]{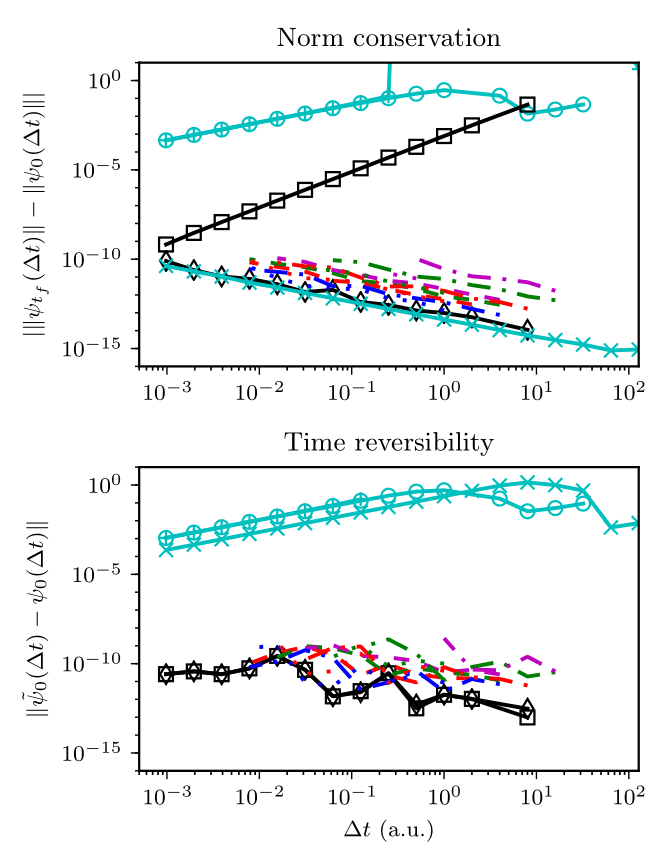}
\caption{Same as Fig~\ref{fig:geometric_properties_nonacs_el_states}, but for energy control in the absence of nonadiabatic couplings.}\label{fig:geometric_properties_nonacs_energy}%

\end{figure}

\begin{figure}
[htb!]\centering
\includegraphics[width=0.5\linewidth]{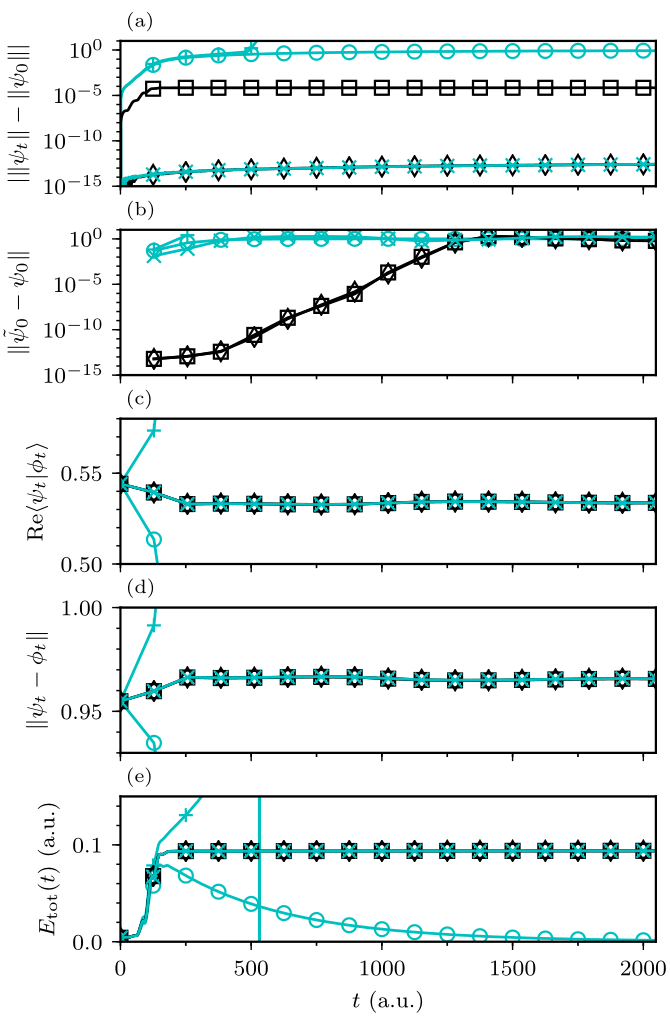}
\caption{Geometric properties of various integrators used for the local population control
in the absence of nonadiabatic couplings.
Panel~(a) shows that only the implicit midpoint and approximate explicit split-operator methods conserve
the norm, while panel~(b) demonstrates that only the implicit midpoint {method and the trapezoidal rule} are
time-reversible. (Reversibility is measured as in Fig.~\ref{fig:geometric_properties_nonacs_el_states}).
Bottom three panels show that no method conserves (c) the inner product, (d) distance
between two states (which would imply stability), or (e) total energy $E_{\text{tot}}(t):=E_{0}(t) +
\langle\hat{\mathbf{V}}_{\text{LCT}}(\psi_{t})\rangle_{\psi_{t}}$
because even the exact evolution operator does not preserve these properties.
State $\phi_{0}$ is $\psi_{0}$ displaced along the reaction coordinate, i.e., a Gaussian wavepacket
with parameters $q_{0}=(0.1,0)$, $p_{0}=(0,0)$, and $\sigma_{0}= (0.128,0)$ a.u.
The time step $\Delta t= 2^{-2}$~a.u. was used for all calculations and line labels are
the same as in Fig.~\ref{fig:error_vs_dt_nonacs_el_states}.
Note that only a few points of the Euler methods are visible in some of the plots because the results of
the Euler methods leave the range of these plots very rapidly.}
\label{fig:nonconserved_geometric_properties_nonacs_el_states}
\end{figure}

\begin{figure}
[htb!]\centering
\includegraphics[width=0.5\linewidth]{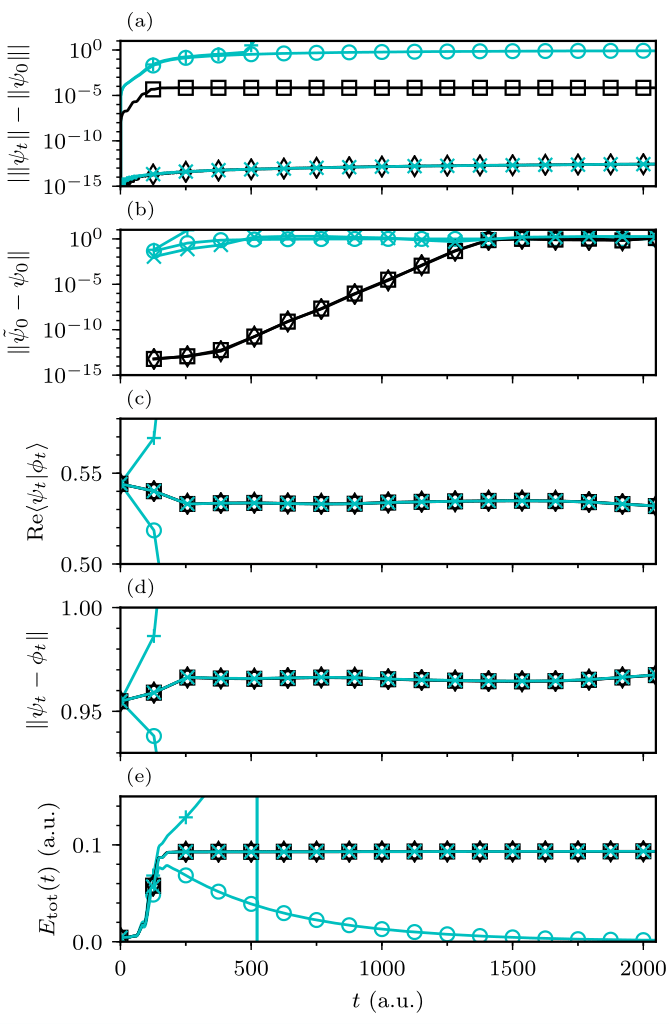}
\caption{Same as Fig. \ref{fig:nonconserved_geometric_properties_nonacs_el_states}, but for energy control with nonadiabatic couplings.}
\label{fig:nonconserved_geometric_properties_nacs_energy}
\end{figure}

\begin{figure}
[htb!]\centering
\includegraphics[width=0.5\linewidth]{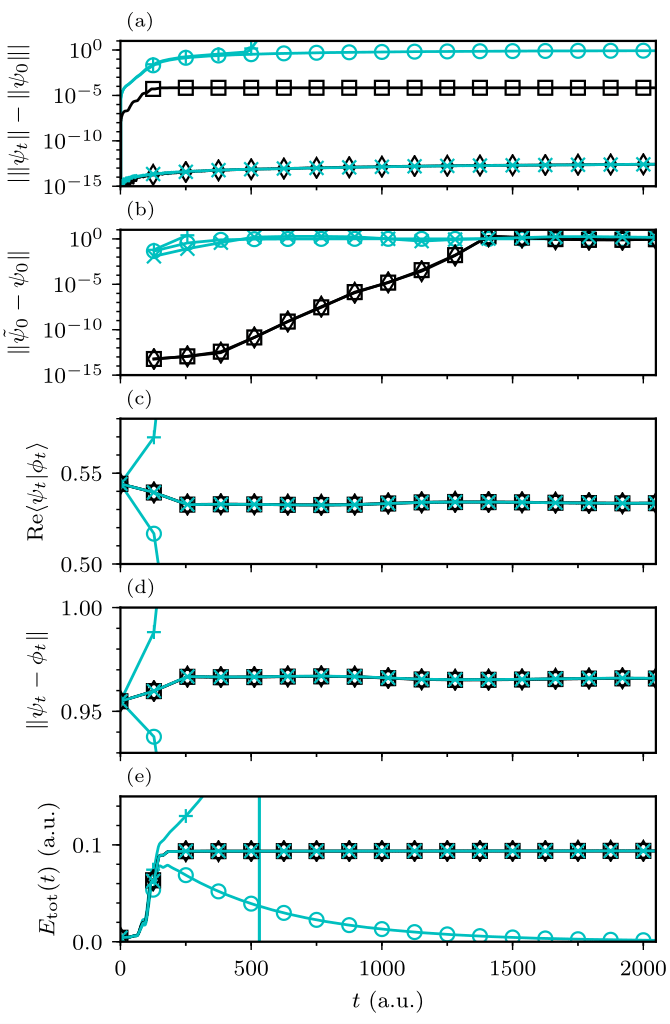}
\caption{Same as Fig. \ref{fig:nonconserved_geometric_properties_nonacs_el_states}, but for energy control without nonadiabatic couplings.}
\label{fig:nonconserved_geometric_properties_nonacs_energy}
\end{figure}

\section{Loss of time reversibility due to roundoff errors}

\begin{figure}
[htb!]\centering\includegraphics[width=0.5\linewidth]{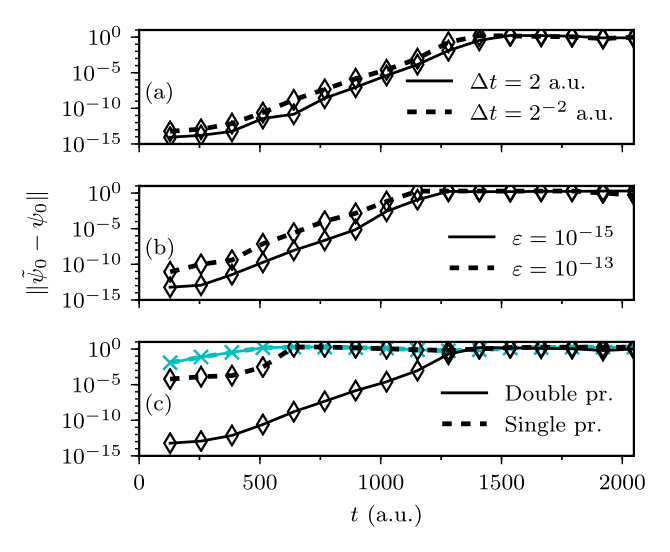}
\caption{Influence of various parameters on the time reversibility of the implicit midpoint method.
The three panels show that the time reversibility of the implicit midpoint method deteriorates
(a) if the time step $\Delta t$ is decreased, leading to a faster accumulation of roundoff errors,
(b) if the error $\varepsilon$ threshold of the nonlinear solver is increased, and
(c) if the precision of the floating-point arithmetic is decreased.
Note that decreasing the floating-point precision does not have a visible influence on the approximate
explicit split-operator algorithm, which is irreversible by construction.
Time reversibility is measured as in Fig.~\ref{fig:geometric_properties_nonacs_el_states}.}
\label{fig:influence_of_parameters_on_tr}
\end{figure}

Figure \ref{fig:influence_of_parameters_on_tr} displays, for the population
control in the presence of nonadiabatic couplings, the influence of various
parameters on the time reversibility of the implicit midpoint method. This
figure aims at demonstrating that the loss of time reversibility at longer times is due to the combination of nonlinearity
and accumulation of roundoff errors and errors due to the
approximate solution of the nonlinear system described by Eq.~(30) of the main
text. Indeed, when using the implicit midpoint method, this
nonlinear system is only solved up to a limited accuracy
$\varepsilon$ (i.e., the nonlinear system is considered solved if $\Vert
f(\psi^{(k+1)})\Vert\leq\varepsilon$).

Panel~(a) shows the influence of the time step on the time
reversibility of the implicit midpoint method. For this, we propagated the
wavefunction to the final time $t_{f}=2048$~a.u. with different time steps and
kept the same error threshold $\varepsilon$ for solving the nonlinear system.
We observe that decreasing the time step decreases the time
reversibility of the implicit midpoint method. This is due to a faster
accumulation of roundoff errors caused by the greater number of steps
performed when using a smaller time step.

Panel~(b) depicts the influence of the error threshold
$\varepsilon$ on the time reversibility. Here, we propagated the wavefunction
using the implicit midpoint method using the same parameters but only changed
the error threshold $\varepsilon$. As expected, a larger error threshold
results in a faster loss of time reversibility as the time increases.

Finally, panel (c) shows the influence of the floating-point
format precision. We compare the single and double precision floating-point
formats for the implicit midpoint method and the approximate explicit TVT
split-operator algorithm. The results indicate that decreasing the
floating-point format from double to single precision, while keeping other
parameters fixed, decreases the time reversibility of the implicit midpoint
method while it does not affect the approximate explicit TVT split-operator
algorithm. Because the implicit midpoint method is by construction time-reversible, decreasing the floating-point
format has the same effect as increasing the threshold $\varepsilon$, namely
increasing the roundoff error when solving the nonlinear system, hence
decreasing its time reversibility. In contrast, the approximate explicit TVT
split-operator algorithm is not affected by the change of precision because it
is, by construction, time-irreversible and would
remain irreversible even if infinite precision were available.

\FloatBarrier
\bibliographystyle{aipnum4-1}
\bibliography{LCT_arbitrary_order}